\documentstyle{elsart}  
\begin{document}
\input psfig.sty
\hyphenation{cal-o-rim-e-ter}

\begin{frontmatter}
\title{Cosmic muon tomography of pure \\ 
cesium iodide calorimeter crystals}

\author[UVa]{E. Frle\v z\thanksref{author}},
\thanks[author]{Corresponding author; Tel: +1--804--924--6786, 
fax: +1--804--924--4576, e--mail: frlez@virginia. edu\hfill}
\author[IRB]{I. Supek},
\author[UVa,Ham]{K. A. Assamagan}, 
\author[PSI]{Ch. Br\"onnimann},
\author[PSI,ETH]{Th. Fl\"ugel},
\author[PSI,DESY]{B. Krause},
\author[ASU,UMas]{D. W. Lawrence},
\author[TSU]{D. Mzavia},
\author[UVa]{D. Po\v cani\'c},
\author[PSI]{D. Renker},
\author[UVa,PSI]{S. Ritt},
\author[UVa,JPL]{P. L. Slocum},
\author[IRB,PSI]{N. Soi\'c}
\address[UVa]{Department of Physics, University of Virginia, 
Charlottesville, VA~22901, USA}
\address[IRB]{Institute Rudjer Bo\v skovi\'c, Bijeni\v cka 46, 
HR-10000 Zagreb, Croatia}
\address[Ham]{Physics Department, Hampton University, Hampton, VA~23668,
USA}
\address[PSI]{Paul Scherrer Institut, Villigen PSI, CH-5232, Switzerland}
\address[ETH]{Institut f\"ur Teilchenphysik, Eidgen\"ossische Technische 
Hochschule Z\"urich, CH-8093 Z\"urich, Switzerland}
\address[DESY]{Deutsches Elektronen Synchrotron, D-22603 Hamburg, Germany}
\address[ASU]{Department of Physics and Astronomy, Arizona State University, 
Tempe, AZ~85281, USA}
\address[UMas]{Department of Physics, University of Massachusetts, Amherst, 
MA~01003, USA}
\address[TSU]{Institute for High Energy Physics, 
Tbilisi State University, 380086 Tbilisi, Georgia}
\address[JPL]{Jet Propulsion Laboratory, Pasadena, CA~91109, USA}

\begin{abstract}
Scintillation properties of pure CsI crystals used in the shower 
calorimeter being built for precise determination of the 
$\pi^+$$\rightarrow$$\pi^0e^+\nu_e$\/ decay rate are reported. 
Seventy-four individual crystals, polished and wrapped in Teflon foil,
were examined in a multiwire drift chamber system specially 
designed for transmission cosmic muon tomography. Critical elements of 
the apparatus and reconstruction algorithms enabling measurement of 
spatial detector optical nonuniformities are described. Results are 
compared with a Monte Carlo simulation of the light response of an ideal 
detector. The deduced optical nonuniformity contributions to the FWHM 
energy resolution of the PIBETA CsI calorimeter for 
the $\pi^+$$\rightarrow$$e^+\nu$\/ 69.8$\,$MeV positrons and the monoenergetic
70.8$\,$MeV photons were 2.7$\,$\% and 3.7$\,$\%, respectively. The upper limit 
of optical nonuniformity correction to the 69.8$\,$MeV positron low-energy 
tail between 5$\,$MeV and 55$\,$MeV was $+\,$0.2$\,$\%, as opposed to the 
$+\,$0.3$\,$\% tail 
contribution for the photon of the equivalent total energy. Imposing the 5 
MeV calorimeter veto cut to suppress the electromagnetic losses, 
{\tt GEANT}-evaluated positron and photon lineshape tail fractions 
summed over all above-threshold ADCs were found to be 
2.36$\pm\,$0.05(stat)$\pm\,$0.20(sys)$\,$\% and 
4.68$\pm\,$0.07(stat)$\pm\,$0.20(sys)$\,$\%, respectively.
\par\noindent\hbox{\ }\par\noindent
\par\noindent\hbox{\ }\par\noindent
\end{abstract}
\end{frontmatter}
\vfill\eject

\section{Introduction}
The PIBETA collaboration has proposed an experimental 
program~\cite{Poc88} with the aim of making a precise determination of 
the $\pi^+$$\rightarrow$$\pi^0e^+\nu_e$\/ decay rate at the Paul Scherrer 
Institute (PSI). The proposed technique is designed to achieve an 
overall level of accuracy in the range of $\sim\,$0.5$\,$\%, improving thus on 
the present branching ratio uncertainty of  4$\,$\%~\cite{McF85}. 
The (1.025$\pm\,$0.034)$\times$$10^{-8}$ pion beta decay ($\pi\beta$) 
branching ratio will be remeasured relative to 
the $10^{4}$ times more probable $\pi^+$$\rightarrow$$e^+\nu_e$\/ decay 
rate, that is known with the combined statistical and systematic 
uncertainty of $\sim\,$0.40$\,$\%~\cite{Cza93,Bri94}. 

The Standard Model description of the $\pi\beta$\/ decay and its radiative 
corrections~\cite{Sir78} enables a stringent test of 
the conserved vector current (CVC) hypothesis and the unitarity of 
the Cabibbo-Kobayashi-Maskawa (CKM) quark mixing matrix. 
The most accurate extraction of the CKM matrix element $V_{ud}$, based
on the superallowed Fermi transitions in nuclei, involves
the theoretical nuclear overlap corrections that dominate
the 0.3$\,$\% total uncertainty~\cite{Sir89}. 
Recent measurements of nine different nuclear transition rates 
confirm the CVC hypothesis at the level of 4$\times$$10^{-4}$, but 
violate the three-generation CKM unitarity condition by more than twice 
the estimated error~\cite{Tow95b}. Complementary neutron decay 
experiments yield a result that also differs from unitarity by over two 
standard deviations, but in the opposite sense~\cite{Tow95b}. It is of 
fundamental interest to measure the branching ratio of the 
$\pi^+$$\rightarrow$$\pi^0e^+\nu_e$\/ decay at the 0.5$\,$\% level or better
to check the consistency with the nuclear and neutron $\beta$\/ decay 
values and the unitary prediction of the Minimal Standard Model 
(MSM). Moreover, the precisely measured $\pi\beta$\/ branching ratio could 
be used to constrain masses and couplings of 
additional neutral gauge bosons in grand unified theories. 
If such couplings are directly discovered at future collider experiments,
precision low-energy data will be an essential ingredient in 
extending the MSM framework~\cite{Mar87}. 

The central part of the PIBETA detector system is a high-resolution,
highly segmented fast shower calorimeter surrounding the active 
stopping target in a near-spherical geometry. 
The requirements imposed on the calorimeter design were:

\begin{itemize}
\item high energy resolution for effective suppression of
background processes;
\item high segmentation and fast timing response 
to handle the high event rates;
\item acceptable radiation, mechanical and chemical 
resistance;
\item compact geometry to simplify the operation and reduce the cost.
\end{itemize}

These design constraints are met in a spherical detector with 
individual calorimeter modules made from undoped cesium iodide (CsI) 
crystals. The nonactivated alkali iodides have been known for more than 
thirty years to exhibit a near-ultraviolet emission component when 
excited by ionizing radiation~\cite{Mur65,Bat77}.
The pure CsI material was reintroduced as a fast, rugged, and relatively 
inexpensive scintillator material by Kobayashi et al.~\cite{Kob87}
and Kubota et at.~\cite{Kub88a,Kub88b}.  
The scintillation characteristics of pure CsI crystals have been
reported subsequently by a number of experimental groups. 
These investigations have covered light readout techniques, 
scintillation decay times, origins of different emission components, 
crystal light yields, and energy and timing 
resolutions~\cite{Kes91,Woo90,Sch90,Utt90,Gek90,Ham95}, 
radiation resistances~\cite{Woo92,Wei93}, wrapping and tuning methods 
and the uniformity of light responses~\cite{Bro95,Dah96}.
The design and performance of pure CsI electromagnetic calorimeters 
have been recently described in Refs.~\cite{Mor89} (NMS at LAMPF/BNL),
\cite{Ray94,Kess96} (KTeV at Fermilab), and~\cite{Dav94} (PHENIX at 
RHIC).

The PIBETA calorimeter geometry shown in Fig.~\ref{fig:ball}
is obtained by the class II geodesic triangulation of 
an icosahedron~\cite{Ken76}. Selected geodesic breakdown 
results in 220 truncated hexagonal, pentagonal and trapezial pyramids 
covering the total solid angle of 0.77$\times$$4\pi$\/ sr. Additional 20 
crystals cover two detector hemispheres open to the beam and act as 
shower leakage vetoes. The inner radius of the crystal ball is 26$\,$cm, 
and the axial module length is 22$\,$cm, corresponding to 12 CsI radiation 
lengths ($X_0$=1.85$\,$cm~\cite{PGP}). There are nine different module 
shapes: four irregular hexagonal truncated pyramids (we label them 
HEX--A, HEX--B, HEX--C, and HEX--D), one regular pentagonal (PENT) and 
two irregular half-hexagonal truncated pyramids (HEX--D1 and HEX--D2), 
and two trapezohedrons which function as vetoes (VET--1, VET--2).
The volumes of our CsI crystal detector modules vary from 797$\,$cm$^3$ 
(HEX--D1/2) to 1718$\,$cm$^3$ (HEX--C). Dimensioned drawings of the
HEX--A and HEX--D1 crystal shapes are shown as examples in two panels of 
Fig.~\ref{fig:shape}.

All key components of the complete PIBETA detector have been prototyped 
and built and have met the required specifications~\cite{Ass95}.
All the detector components were delivered to PSI by 
mid-1998. The final assembly of the apparatus was followed by the in-beam 
commissioning and calibration. 
``Production'' measurements are scheduled for 1999, and will extend over
two more years to complete the first phase of the project, 
a $\sim\,$0.5$\,$\% measurement of the $\pi^+$$\rightarrow$$\pi^0e^+\nu$\/ 
decay rate. 

The quality of the delivered CsI calorimeter crystals was controlled in
the following program of measurements:
\begin{enumerate}
\item check precisely the crystal physical dimensions against the 
      specifications;
\item study the effects of different crystal surface treatments and 
      wrapping configurations on the energy and timing 
      resolution and on the uniformity of light response;
\item determine the percentage of the fast scintillation component in the
light output;
\item measure the contribution of detector photoelectron statistics to 
energy resolution;
\item find the temperature dependence of a detector's ADC readings;
\item determine axial and transverse optical nonuniformities for each
crystal.
\end{enumerate}

The ultimate goal of the program was to provide the input parameters 
for a Monte Carlo simulation of the calorimeter response and to deduce 
the effects of CsI crystal light yields and optical nonuniformities 
on the energy lineshapes of detected photons and positrons ranging in
energy between 10$\,$MeV and 120$\,$MeV.

The spectra generated by cosmic muons have been used to examine 
the uniformity of light response of two large (24$\times$36$\,$cm) 
cylindrical NaI(Tl) calorimeter crystals in the earlier work of Dowell 
et al.~\cite{Dow90}. 
Their apparatus relied on plastic scintillator hodoscopes with 
3.8$\times$3.8$\,$cm$^2$ cross section for the charged particle tracking. 
The precision of the trajectory reconstruction was therefore limited to
$\sim\,$4$\,$cm. Such positional resolution was deemed inadequate for our 
application because the PIBETA CsI crystals have an axial length of 
22$\,$cm with front (back) surface side lengths of less than 4$\,$cm 
(7$\,$cm).

Most of our measurements were done with the tomography apparatus built 
around three pairs of multiwire drift chambers (MWDC) using cosmic 
muons as the probe. One dozen PIBETA CsI detectors were examined in 
a 350$\,$MeV/c minimum ionizing muon beam of the PSI $\pi$E1 area, as well 
as in the 405$\,$MeV/c penetrating pion and stopping proton beams in 
the $\pi$M1 channel. Selected CsI crystals were also scanned with a 662 
keV $^{137}$Cs $\gamma$-source. The light yield nonuniformities measured
in the beam and with a radioactive-source scan method were compared with
more precise cosmic muon tomography data. The experimental apparatus
used for a $\gamma$-source crystal scans and the subsequent data
reduction were simple enough to be easily adopted by the crystal
manufacturers, thus accelerating the subsequent quality control cycle.

\section{Crystal Production, Mechanical Quality Control and Surface
Treatment}
Pure CsI crystals produced for the PIBETA calorimeter came from two 
different sources. Twenty-five crystals were grown and cut to specified
shapes in the Bicron Corporation facility in Newbury, Ohio. 
The rest of the scintillators were grown in the Institute for
Single Crystals in Kharkov (AMCRYS), Ukraine. 
Preliminary examination of fifty-five AMCRYS CsI crystals, including 
the mechanical and physical tests, was done in the High Energy Institute 
of Tbilisi University, Georgia.

Manufacturing tolerances of the crystals were specified for the linear 
dimensions ($+$150 $\mu$m/$-$50 $\mu$m) and for the angular deviations 
($+$0.040$^\circ$/$-$0.013$^\circ$). Geometrical dimensions of the 
machined crystals were measured upon delivery at PSI 
using the computer-controlled distance-measuring device {\sl WENZEL
Precision}. The machine was programmed to automatically probe the 
surfaces of a subject crystal with a predefined shape. Each crystal surface 
plane was scanned with a touch head at six points and the equations of planes 
were found through these surveyed points. Body vertices, measured with 
an absolute precision of 2 $\mu$m and reproducible within 20 $\mu$m, 
were then compared with the expected theoretical values. Those crystals 
that failed the imposed geometrical tolerances were returned to 
the manufacturer for reuse as raw crystal growing material. 

After physical measurements crystal surfaces were polished with a mixture 
of 0.2 $\mu$m aluminum oxide powder and etylenglycol. Next came a measurement 
of the light output of the fast CsI scintillation component (F) that is
completely decaying in the first 100 ns, relative to the total CsI signal 
(T), integrated in a 1 $\mu$s ADC gate. These measurements
were made with unwrapped crystals using air-gap coupled 
photomultiplier tubes (PMTs) and a Tektronix TDS 744 digital 
oscilloscope. Only the crystals with a fast-to-total component ratio 
(F/T) better than 0.7 were accepted. The mean value 
of fast-to-total ratio for all accepted CsI crystals was 0.788,
as shown in Fig.~\ref{fig:ft}.

EMI photomultiplier tubes 9821QKB~\cite{EMI} with 75$\,$mm diameter cathode
were glued to the back faces of hexagonal and pentagonal CsI crystals using 
a 300 $\mu$m layer of silicone Sylgard 184 elastomer (Dow Corning RTV silicon 
rubber plus catalyst). Smaller half-hexagonal and trapezial detector 
modules were equipped with two inch EMI 9211QKA phototubes~\cite{EMI}. 
The resulting crystal--photomultiplier tube couplings were strong and 
permanent, but could be broken by application of a substantial 
tangential force. The PMT quartz window transparency, peaking 
at $\sim\,$380 nm~\cite{EMI}, is approximately matched to the spectral
excitation of a pure CsI fast scintillation light component with 
a maximum room temperature emission at $\sim\,$310 nm~\cite{Woo90}. 
The PMT high voltage dividers were modified EMI-recommended bases 
designed and built at the University of Virginia. These dividers 
minimized the so-called ``super''-linearity exhibited by many PMTs 
well below the onset of saturation. The maximum PMT nonlinearity 
measured with a pair of light-emitting diodes was less than 2$\,$\% 
over the full dynamic range expected in the $\pi\beta$\/ decay rate 
measurement~\cite{Col95}.

Four different wrapping materials for lateral 
crystal surfaces were investigated in wrapping and tuning studies with 
405$\,$MeV/c pion and proton beams in PSI $\mu$M1 area:
\begin{enumerate}
\item one to five layers of a 38$\,$$\mu$m PTFE Teflon sheet 
(CF$_2$\/ monomer, $X_0$=16.0$\,$cm);
\item one to five layers of 30$\,$$\mu$m Mylar sheet (C$_5$H$_4$O$_2$, 
      $X_0$=28.7$\,$cm);
\item a 250 $\mu$m thick Tyvek fleece (polyethylene CH$_2$, 
$X_0$=131.3$\,$cm);
\item a 110$\,$$\mu$m thick Millipore filter (polyvinylidene fluoride PVDP,
$X_0$=19.0$\,$cm) with 0.22$\,$$\mu$m pores;
\item a wavelength shifter lacquer treatment of crystal surfaces.
\end{enumerate}
 
In all cases a primary diffuse reflector was protected by an additional 
20 $\mu$m thick aluminized Mylar cover. These preliminary tests showed
more uniform axial light collection from the crystal front section when 
a polished front crystal surface was covered with a black paper sheet. 
A similar tuning method, using black paper strips at the front section
of 5$\times$6$\times$36$\,$cm$^3$ CsI(Tl) crystals in order to improve 
the uniformity of light collection, was recently described in
Ref.~\cite{Dah96}.

The maximum fast-to-total ratio was always achieved with unwrapped 
crystals because all applied wrapping materials absorb more ultraviolet
light which dominates the fast component, than visible light.
A two-layered Teflon cover was found to be superior in delivering 
$\ge$20$\,$\% more fast scintillation light component than the other 
reflectors. It was also comparatively better in not degrading the F/T 
ratio by more than $\sim\,$2$\,$\%. These differences in measured light output
were reproducible in the beam tests with $\sim\,$1$\,$\% event statistics and
estimated 2$\,$\% systematic uncertainty. They are subsequently confirmed 
in more controlled cosmic muon tomography measurements. 

On the basis of these findings the adopted standard wrapping configuration 
for all studied CsI crystals consisted of: (a) the lateral surfaces being 
wrapped with two layers of Teflon foil plus one layer of aluminized Mylar 
sheet, and (b) the front crystal surface covered with the black paper 
template. The back crystal surface with the glued phototube was left uncovered.

Light yield measurements were repeated for a subset of crystals after 
a six month period to confirm that no appreciable changes occurred
as a result of degradation of the surface reflectivities and wrapper 
material qualities. The measured light output was typically within 5$\,$\% 
of the originally measured value.

The most successful surface treatment of CsI crystals, however, involved 
painting the lateral crystal surfaces with a waveshift lacquer. We
recently studied this method on a large sample of CsI detectors.
The light response and uniformity properties were noticeably improved,
resulting in $\sim\,$20$\,$\% better energy resolutions for the 70$\,$MeV 
monoenergetic $e^\pm$ and 50--82$\,$MeV tagged photons. The possible 
degradation and change in the crystal surface reflectivities that could be 
difficult to account for in the multi-year long precision experiment like 
the PIBETA is also expected to be arrested by such a treatment. These 
measurements will be covered in detail in a forthcoming paper~\cite{Frl98}.

\section{Tomographic Apparatus}
A simplified sketch of the tomographic apparatus layout is illustrated in 
a {\tt GEANT} rendering showing a few simulated cosmic muon 
trajectories on Fig.~\ref{fig:drift}. Three identical delay-line readout 
drift chambers were used to define the cosmic muon tracks intersecting CsI 
crystals. The chambers were built at
the Los Alamos Meson Physics Facility (LAMPF) by the M. Sadler group from
the Abilene Christian University. This type of drift chambers has been 
used reliably for years with several LAMPF 
spectrometers~\cite{Ate81,Mor82,Ran81,Ran82}. Each chamber 
consists of two signal and three ground planes with the nominal 
active area of 60$\times$60$\,$cm$^2$. The horizontally oriented ``$x$'' 
and ``$y$'' signal planes are two orthogonal sets of alternating cathode
and anode wires evenly spaced at 0.4064$\,$cm.

The light-tight aluminum box had two drift chamber pairs mounted on its top 
plate, and one pair fixed below the bottom side. Distances between 
the centers of chamber pairs 1--3 and 2--3 were 24$\,$cm and 27$\,$cm, 
respectively.
The dark box could accommodate up to six CsI calorimeter modules at one 
time. It had feedthrough connectors for twelve signal and twelve high 
voltage cables as well as six temperature sensor lines and six 
LED calibration signal cables. A pair of 1$\,$cm thick plastic 
scintillators were placed directly below the apparatus and separated 
from the chamber system by a 5$\,$cm layer of lead bricks shielding out 
the soft cosmic ray component. The Monte Carlo simulation 
(see Sec.~\ref{mcd})showed that hard cosmic muons penetrating the frames of 
the apparatus, CsI crystals and shielding material and triggering two 
scintillators had a smooth energy threshold starting at $\sim\,$120$\,$MeV. 
The zenithal angle range of accepted cosmic muon trajectories intersecting 
all three MWDCs and at least one CsI detector was $\pm\,$45$^\circ$.

The chambers were operated with anode wires held at positive voltage 
between 2400--2600 V. The chamber cathode wires were grounded. The gas
mixture was 65$\,$\% argon, 35$\,$\% isobutane plus 0.1$\,$\% isopropyl 
alcohol. Detection efficiencies of individual chambers for penetrating cosmic 
muons exceeded 90$\,$\%, with the combined six-chamber efficiency routinely 
surpassing 50$\,$\%.

Anode wires were attached directly to a fast 2.5 ns/cm delay-line.
Signals from both ends of the delay line were amplified twenty-fold, 
discriminated at a threshold of 10 mV and connected to two channels 
of a time-to-digital converter (LCR 2229A TDC). The time difference 
in the two TDC readings identified the fired anode wire.
Cathode lines defining the alternate field~\cite{Wal78,Ers82} were 
connected to one ``odd'' (O) and one ``even'' (E) line, and the signals 
on these lines were processed by a custom-made electronics unit which 
added and subtracted the analog pulses~\cite{Bro79}. The electronic sum 
of cathode pulses (O$+$E) was discriminated and the resulting delayed
signal determined the drift position timing, and was used to stop 
another TDC channel. The difference of the cathode signals (O$-$E) was 
digitized with an analog-to-digital converter (LCR 2249A ADC).
That information was used to discriminate between the events that produced
the ionization tracks left and right of the given anode wire.
The 100 ns integration gate timing was defined by the (O+E) logic pulse
timing. Four measurements were therefore required to find 
an intersection between a cosmic muon track and one drift chamber plane, 
namely three TDC values and one ADC value.
The schematic diagram of the electronic logic is shown in 
Fig.~\ref{fig:tmelec1}.

Advantages of the system were good charged particle track resolution 
($\sim\,$0.5$\,$mm root-mean-square in the horizontal plane, 
Fig.~\ref{fig:deviation}), stability to fluctuations in the outside 
environmental parameters (humidity, temperature and pressure) and low 
cost of the associated electronics logic and readout. Limited counting 
rate and restriction to single hit events in the chambers did not 
represent a drawback in this tomographic application. The apparatus 
was operated in an air-conditioned room with a controlled humidity 
level kept below 30$\,$\%. The temperature at six points inside the dark 
box as well as absolute time were recorded for every triggered event. 
The temperature range recorded inside the dark box during three years 
of data taking was (22$^\circ$$\pm\,$3$^\circ\,$)C. The typical temperature 
gradient was 0.4$^\circ\,$C/day, leading to average absolute temperature 
variation of  0.2$^\circ\,$C and average CsI ADC gain drift of 
$\sim\,$0.4$\,$\% 
in a single data taking run (fixed at 250000 triggers, $\sim\,$6 hours). 
Measured light output variations of the CsI crystals and the PMT gain 
drifts caused by daily temperature cycles were compensated for in 
the replay analysis (see Sec.~\ref{raw}).

\section{Data Acquisition System}\medskip
The computer code used for data acquisition was HIX (Heterogeneous 
Information Exchange), a data acquisition system developed by S. Ritt 
at PSI, originally intended for use in small and medium-size nuclear 
physics experiments~\cite{Rit95}.

The ``frontend'' 486 personal computer was connected to a CAMAC
crate via a HYTEC 1331 interface that read ADC, TDC, scalar, and 
temperature sensor units. The C program running under MS-DOS accessed 
data and sent it over the network to a VAX 3100 server. A simple
communication program on a VAXstation received data from the frontend 
computer and stored it in a global section buffer. Buffered data were 
passed by the Logger application to a user-written Analyzer program.\
Analyzed events were subject to predefined cuts, filled in predefined 
histograms, and stored in a raw data stream directed 
simultaneously to a hard disk drive and an 8-mm tape system. 
The experiment was controlled from a PC computer by a Microsoft Windows 
control program. A windows-based graphical user interface to the program
allowed starting, pausing and stopping data acquisition, as well as
online inspection of individual raw data words, calculated data 
words, scalers, assorted efficiencies, one and two-dimensional 
histograms, and gate, box and Boolean tests. Different PC computers 
on the Ethernet could make a connection to the VAX server called Link 
and access the same information remotely.

During data acquisition only those events for which at least two $x$\/ and 
two $y$\/ drift chambers had nonzero ADC and TDC values were written to
an disk-resident ASCII Data Summary File (DSF). One good event
contained 55 raw data words specifying the response of the MWDCs and CsI
detectors, instantaneous temperatures and absolute times. That was 
a full set of observables for which a cosmic muon trajectory could be 
unambiguously reconstructed. The raw trigger rate was $\sim\,$13 Hz with 
six CsI crystals in the dark box. The average DSF event rate was 
$\sim\,$4 Hz. Individual runs were stopped and restarted automatically
after 250,000 collected triggers. A total of about 10$^5$ DSF events 
per crystal were typically collected in one week of data acquisition. 
Tomography data for one set of six crystals were usually collected over
a two week period.

\section{Drift Chamber Calibration}
Two different trigger configurations were used interchangeably during 
the data collection. Drift chamber calibration was done with a trigger
requiring a two-scintillator coincidence and good ADC and TDC data for
at least one drift chamber. In the tomography data acquisition mode, 
the triple coincidence between two tag scintillators and one CsI 
detector was required, accompanied by at least one good pair of $x$\/ and 
$y$\/ chamber hits. The rare accidental coincidences involving signals in 
more than one CsI detector were eliminated from the DSF records in 
the offline analysis.

The hit anode wire number $n_A$\/ was calculated from
\begin{equation}
n_A={{(T_1-T_2+N\cdot D)}\over {2D}},
\end{equation}
where $T_1$\/ and $T_2$\/ were the TDC values for two ends of the anode delay
line, $N$\/ was the number of wires in a chamber (between 71 and 73), and
$D$\/ was the $\sim\,$2.05 ns time delay between the adjacent wires.
Truncated wire position $x_T$\/ was defined as the nearest integer multiple
$n_A$\/ of the 0.8128$\,$cm wire separation $g$:
\begin{equation}
x_T={\tt NINT}(n_A\cdot g).
\end{equation}

The anode delay time difference $(T_1-T_2)$\/ depended nonlinearly on the
anode position. Linearized anode wire positions for each chamber were 
approximated by a polynomial expansion
\begin{equation}
x_A=\sum_{i=0}^m c_i\cdot (T_1-T_2)^i,
\end{equation}
where the order $m$\/ of the fit with $\chi^2$ per degree of freedom
$\chi^2/(m-1)$$\approx\,$1 depended 
on the particular delay line. For our chambers the orders of the fits
were 3, 4, or 5. The coefficients $c_i$\/ were optimized with the program 
{\tt MINUIT}~\cite{Jam89} that minimized the sum of squared 
deviations $(x_T-x_A)^2$\/ for each chamber in every run.

The drift distance $d_x$\/ was determined from the drift timing $T_x$\/ using
the calibration lookup table $f(T_x)$
\begin{equation}
d_x=f(T_x)\cdot T_x+\Delta T_{x},
\end{equation}
where $\Delta T_{x}$\/ was the drift time offset. The lookup table was 
calculated assuming that incoming cosmic muons are distributed uniformly
over the equidistantly spaced wires.

The final hit position $x$\/ was given by expression:
\begin{equation}
x=x_T+(-1)^n\cdot sign({\rm ADC_{O-E}+ADC_0})\cdot d_x+x_{\rm OFF},
\end{equation}
where $x_{\rm OFF}$\/ was the chamber $x$\/ coordinate offset, and
${\rm ADC_{O-E}}$\/ and ${\rm ADC_{0}}$\/ were the (O$-$E) ADC
signal and associated offset, respectively.

The individual chamber phases $n$\/ in the Eq. (5) and absolute horizontal 
coordinate offsets $x_{\rm OFF}$ and $y_{\rm OFF}$ between six chambers were 
adjusted with help of two-dimensional histograms 
$(x_1-x_3)$\/ vs $(x_2-x_3)$\/ and 
$(y_1-y_3)$\/ vs $(y_2-y_3)$~\cite{McN87,Sup89}.

\section{Raw Data Reduction}\label{raw}

Data summary files were analyzed offline by compressing the DSF event data
into {\tt PAW} Ntuples~\cite{Bru93}. Drift time-to-distance lookup tables
and assorted chamber calibration constants, as well as CsI detector ADC 
pedestals and ADC temperature corrections, were determined for each run 
separately. The high voltages of the CsI detector phototubes were selected to 
give $\sim\,$8 ADC channels/MeV scale. The FWHMs of the pedestal peaks
were typically 2 ADC channels. Therefore, the pedestal widths 
corresponded to an energy deposition of 0.25$\,$MeV, 24 
$\mu$m pathlength or 0.5$\,$\% of the most probable energy deposition by 
a minimum ionizing charged particle. 
Drifts in the pedestal position over a one month period amounted to less 
than 2 ADC channels.

The precise horizontal coordinate offsets of six CsI crystals inside
the dark box were determined in the next stage of analysis. 
Using the preliminary offsets read from the plastic template on which
the crystals were laying, 
between 5 and 10 percent of the reconstructed cosmic muon trajectories 
were found not to intersect any of the predefined detector volumes. After 
adjusting the $x$\/ and $y$\/ translation software offset parameters 
of the CsI modules by maximizing the number of tracks intersecting individual
crystal volumes with nonzero ADCs, the real number of events undergoing 
scattering in the apparatus or having the improperly reconstructed tracks 
was shown to be below 1$\,$\%. That percentage was in agreement with the Monte 
Carlo simulation of cosmic muons interacting with the experimental apparatus, 
showing the most probable muon scattering angle of 0.25$^\circ$ and 
the 2.5$^\circ$ mean scattering angle. The Monte Carlo root-mean-square of 
the scattering angle for the accepted events was 0.66$^\circ$, 
Fig.~\ref{fig:ths}. That value translates into an average pathlength 
uncertainty of 1$\,$mm. The estimated error in finding the correct crystal 
position inside the dark box was smaller than $\sim\,$0.5$\,$mm, 
Fig.~\ref{fig:position22}.

The energy deposited in CsI crystal by cosmic muons along the fixed
pathlength is a broadened distribution due to the statistical nature of
energy transfers. Maximal measured pathlengths were up to 12$\,$cm, while the 
average pathlength was $\sim\,$6$\,$cm. A Gaussian model for the energy loss 
distribution of the minimum ionizing particles in CsI thicknesses of
less than 12$\,$cm is not a good approximation. The distribution of energy
losses is asymmetric, with a long high energy tail; the most 
probable energy loss is smaller than mean energy loss~\cite{Leo87}.

Light yield temperature coefficients for the individual CsI 
detectors were determined by the ``robust'' estimation~\cite{Hub81}
of the ADC values per unit pathlength as a linear function of 
temperature recorded inside
the dark box. The least-squares condition assumes normally
distributed measurements with constant standard deviations and is 
therefore not appropriate in this application, as pointed out above.
Typically, more than $10^5$ cosmic muon events, collected over 
at least one week and spanning the temperature range 
of $\sim\,$2$^\circ$C, were used as an input experimental data set. 
The {\tt FORTRAN} subroutine {\tt MEDFIT}, documented in 
the Ref.~\cite{Pre86}, was adopted as the fitting procedure by imposing
the requirement of minimum absolute deviation between the measured
and calculated ADC values per unit path that were dependent linearly on
the temperature variable.
The average light output temperature coefficient for 74 different CsI 
detectors extracted by that method was $-1.4\pm\,1.4$$\,$\%/$^\circ$C, both 
for fast and total scintillation light components. 
These numbers are in good agreement with the previously reported value of 
$-$1.5$\,$\%/$^\circ$C~\cite{Kob87}, but the spread of extracted 
coefficients for different CsI detectors was large: $\pm\,$4$\,$\%/$^\circ$C 
(Fig.~\ref{fig:tf}). We point out that the method did not allow us to 
separate the temperature dependence of the crystal light outputs from the 
temperature instabilities of the phototubes and high voltage dividers 
as well as the temperature instabilities of the ADC modules themselves. 
The LeCroy catalogue specification~\cite{LCR95} lists the typical 
temperature coefficient of a LCR 2249A ADC unit as zero, and the maximum 
coefficient up to $\pm\,$3$\,$\% for an ADC gate longer than 100 ns and 
an average ADC reading of 75 pC (about a quarter of a full 256 pC scale). 

The raw ADC values were corrected for the temporal 
temperature variations and written into the revised DSF files used in 
subsequent analysis.

\section{Detector Photoelectron Statistics}

The contribution of the photoelectron statistics to the energy resolution 
of CsI crystals was determined by a photodiode-based system. Six 
individual detectors (CsI crystals with photomultiplier tubes and high 
voltage dividers) having identical light emitting diodes (LEDs) coupled to 
their back sides were placed inside the cosmic muon tomography apparatus. 
The LEDs were pulsed at a 10 Hz rate using a multichannel diode driver with
continuously adjustable output voltage. One split output signal from 
each channel of the driver generated a 100 ns wide ADC gate. The LED 
light in a CsI detector produced fast ($\sim\,$20 ns FWHM) PMT anode pulses
whose integrated values depended on the driving voltage and were equivalent
to the fixed cosmic muon energy depositions between 10$\,$MeV and 100$\,$MeV.

A total of five different LED amplitudes were used in measurements of 
each CsI detector. Examples of the LED spectra are shown in  
Fig.~\ref{fig:set2}. The ADC pedestals were recorded simultaneously during
the run. The intensity of LED light was cross-calibrated against
the cosmic muon spectra peaks in CsI crystals. These muons were tracked
in tomography drift chambers and both their pathlengths and energy
depositions in the crystals could be easily calculated. That calculation
enabled the establishment of the absolute energy scale in MeV.

The variance $\sigma_E^2$\/ of a photodiode peak depended upon the mean
number of photoelectrons $\bar N_{pe}$\/ on the photocathode created per
unit energy deposition in the crystal:
\begin{equation}
\sigma_E^2=\sum_i\sigma_i^2+\bar E/\bar N_{pe},
\label{eq:led}
\end{equation}
where $\bar E$\/ was the LED spectrum peak position and $\sigma_i^2$'s were 
assorted variances, such as instabilities of the LED driving voltage and 
temporal pedestal variations.
Five measured points ($\sigma_E^2$,$\bar E$) were fitted with a linear 
function (\ref{eq:led}) and the mean number of photoelectrons per MeV
$\bar N_{\rm pe}$\/ for each CsI detector was determined. The statistical error 
of the least-squares fits was typically less than one
photoelecton/MeV. Using the previously extracted light output
temperature coefficients all $\bar N_{\rm pe}$\/ values
were scaled to the 18$^\circ$C point. That value was the designed 
operating temperature of the PIBETA calorimeter.

The number of extracted photoelectrons per MeV for the fast 
scintillation component fell into the 20--130 range as shown in 
Fig.~\ref{fig:npf}. The hexagonal and pentagonal crystals equipped with
three inch phototubes averaged 73 photoelectrons/MeV, while 
the half-hexagonal and veto detectors with two inch phototubes
had a mean of 33 photoelectrons/MeV. The 73/33 ratio is very well
explained by the (3/2)$^2$ ratio of photosensitive areas 
for two different photocathode sizes. The measured photoelectron
statistics were in agreement with the 100 ns ADC gate values of
$\bar N_{\rm pe}$=(20--260) for large ($\sim\,$10$\,$cm) pure CsI crystals 
equipped with two or three inch PMTs
reported in the past in Refs.~\cite{Woo90,Sch90,Utt90,Mor89}.

The quantum efficiency of our bialkali photocathodes for pure CsI 
scintillation light was 23$\,$\%~\cite{EMI}, the average light collection 
probability for our detector  shapes 23$\,$\% (Sec.~\ref{tks}) and the 
fraction of deposited energy converted into the scintillation light was about
12$\,$\%. Therefore, starting from the mean number of 73 photoelectrons/MeV, 
we calculated that 1$\,$MeV energy deposited in the CsI crystal produced on 
average about 10$^4$ scintillation photons.

\section{Monte Carlo Description}
\subsection{Tomography System}\label{mcd}
Geometrical layout of the experimental apparatus was defined using the
{\tt GEANT} detector description and simulation tools~\cite{Bru94}. 
The active elements of the simulated apparatus were an aluminum box
housing six CsI modules, six multiwire drift chambers,
two scintillator planes and a layer of lead brick shielding, all shown in
Fig.~\ref{fig:drift}. Generated events were muons with the energy,
angular and charge distribution of a hard cosmic ray component at sea 
level. The zenithal angle $\theta_z$\/ distribution of muons at 
the ground was assumed to be proportional to $\cos^2\theta_z$;
the momentum spectrum between 0.1 and 10$^3$ GeV/c and 
the energy-dependent ratio of number of positive to number of negative
cosmic muons was parameterized from the data given in 
Refs.~\cite{PGP,Ros48}.

The simulation trigger was defined by requiring a minimum energy 
deposition of more than 0.2$\,$MeV in each scintillator plane. This
threshold was about one-tenth of the minimum-ionizing peak in 
the triggered scintillator and corresponded to the discriminator
level used in the data acquisition electronics. 

Penetrating cosmic muons and generated secondary particles
were tracked through the apparatus, and energy depositions, pathlengths
in CsI crystals and hits in the drift chambers were digitized. All 
physical processes were turned on in the {\tt GEANT} programs with
the cutoff energies of 0.2$\,$MeV for charged particles and photons.
Inspection of simulated energy deposition spectra showed deviations
from theoretical Vavilov distributions~\cite{Leo87} expected in the 
planar detector geometries. The differences were caused by multiple
scattering in the irregularly shaped crystals, and were particularly
prominent for the events with shorter pathlengths close to the crystal
edges. The simulated ADC histograms revealed that triggering cosmic muons
deposit on average 10.33$\,$MeV/cm in a CsI detector, with the most 
probable energy loss 5.92$\,$MeV/cm.

Cosmic muon spectra in CsI detectors were described in a satisfactory way
($\chi^2/(m-1)$$\approx\,$1.2) by the combination of a Gaussian distribution
and a falling exponential tail:
\begin{equation}
{A}(\epsilon)=\theta(\epsilon_0-\epsilon)e^{ -{1\over 2}[{{(
\epsilon-\bar\epsilon})/{\sigma_\epsilon}}]^2 }+\theta(\epsilon-
\epsilon_0)e^{\alpha-\beta\epsilon},
\end{equation}
where $\epsilon$\/ is energy (in MeV) deposited in one full CsI module.
Parameters $\bar\epsilon$, $\sigma_\epsilon$, $\epsilon_0$,
$\alpha$, and $\beta$\/ are extracted from the {\tt GEANT} spectra by
imposing the least squares constraint to the fit and leaving
the pathlength $d$\/ (cm) as free parameter:
\begin{eqnarray}
\bar\epsilon(d)&=&5.079+0.1876d-9.9390\cdot 10^{-3}d^2, \\
\sigma_\epsilon(d)&=&0.3545+7.2329\cdot 10^{-3}d-3.4148\cdot 10^{-4}d^2,
\\
\epsilon_0(d)&=&5.1800+0.1989d-1.0297\cdot 10^{-2}d^2, \\
\alpha(d)&=&4.0581+0.4465d-1.5760\cdot 10^{-2}d^2, \\
\beta(d)&=&0.7952+5.0313\cdot 10^{-1}d-1.5240\cdot 10^{-3}d^2, \ 
\label{eq:mc}
\end{eqnarray}
\centerline{and\hfill}
\begin{eqnarray}
\theta(\epsilon_0-\epsilon)=\cases{ 0,& if $\epsilon_0<0$,\cr
                                     1,& otherwise.\cr }
\end{eqnarray}
These Monte Carlo spectra, broadened with photoelectron statistics,
were used to describe our cosmic muon lineshapes produced by the 
optically uniform CsI detectors.

\subsection{Light Collection Simulation: {\tt TkOptics} Code}\label{tks}
Propagation of scintillating photons through a uniform detector with
ideal reflecting dielectric surfaces and different wrapping
materials was studied using the {\tt TkOptics} simulation 
program~\cite{Wri92,Wri94}. The code is a library of {\tt FORTRAN}
subroutines with an X-Windows-based user interface written with
{\tt Tck/Tk} toolkit~\cite{Ous94}. The program can simulate the light 
output response of an arbitrarily shaped scintillation detector 
with given bulk and surface optical properties. 

The polygonal detector shape is defined by the coordinates of its 
vertices. The detector attributes are reflector types of lateral and
front crystal surfaces and wrapper material, crystal surface-wrapper gap
distances and characteristics of the photomultiplier tube and 
a phototube-crystal joint coupling. The program handles normal
dielectric, specular, diffuse and partially absorbent reflector
surfaces with arbitrary diffuse fraction, roughness, and specular
$r_s$\/ and diffuse reflectivity $r_d$. Predefined bulk properties of 
a detector are the medium scattering and attenuation length as well as
refractive index. A photomultiplier tube is specified by the diameter, 
position and quantum efficiency of the photocathode. 

Different choices of initial scintillating photon distribution are
possible, the default being a uniform starting distribution throughout
the scintillating volume. A working volume is a box divided into
elementary cubic cells of fixed size. Output menu options include
the initial and endpoint photon coordinates, and direction vectors and
timing distributions of the detected scintillation photons organized
in a {\tt PAW} Ntuple. Every elementary cell is flagged as a bulk or
edge crystal cell. Center coordinates of the cells and a fraction
of photons generated in every cell, and subsequently detected on
the PMT sensitive surface, are always recorded. Results of high
statistics runs with $10^7$ photons generated uniformly through
the detector volume were plotted to show the number of photoelectrons 
as a function of scintillation source position inside
the crystal. The relative statistical errors of calculated light
collection probabilities for the bulk crystal cells were $\le\,$2$\,$\%. 

The bulk attenuation length and the light scattering length of the 
near-ultraviolet emission component inside the
CsI crystals were set to 150$\,$cm and 200$\,$cm, respectively~\cite{Frl89}. 
The index of refraction for the CsI medium increases from 1.82 for
the blue-green light to about 2.08 for ultraviolet
light~\cite{Frl89}.

The simulated light collection probabilities $P(x,z)$\/ that depend on
axial $z$ and transverse $x$ positions are shown integrated in the vertical 
coordinate ($y$) in Figs.~\ref{fig:hexa_9} and~\ref{fig:hh1d_9}. 
in the form of ``lego'' plots for one wrapping configuration and two 
different crystal shapes. The coordinate system is defined with 
the $z$=0$\,$cm origin at the front face of the crystal and a photocathode 
window at the $z$=22$\,$cm plane.

We find that simulated light response of the ideal hexagonal or
pentagonal PIBETA CsI detector with specular lateral surfaces and
a diffuse wrapping material can be described with the following parameters:
\begin{enumerate}
\item The average photon collection probability $P$\/ for a three inch
photocathode is about 23$\,$\% for
ideal $r_s$$=$1.0 crystal surfaces and a $r_d$$=$0.9 diffuse wrapper,
decreasing by half, to 12$\,$\% for $r_s$$=0.8$.
\item The axial light collection probability variation through the 
first 10 centimeters is in the $\pm\,5\,$\% range, with a positive slope 
$dP/dz\ge 0$ for higher specular and diffuse reflectivities, namely for 
$r_{s,d}$$\ge$0.9.
\item The axial detected light variation in the back crystal half 
($z$$\ge$10$\,$cm) is up to $-30\,$\%.
\item The transverse light response referenced to the light yield at the
crystal axis typically increases towards the crystal surfaces by up to 
$+\,$5$\,$\% for $z$$\le$10$\,$cm, but is generally declining away from
the central axis by $-\,$30$\,$\% at the $z$=18$\,$cm plane.
\item The root-mean-square of a three-dimensional light nonuniformity 
is between 2.5$\,$\% and 3.5$\,$\% for $z$$\le$10$\,$cm, compared to 
a $\sim\,$20$\,$\% 
root-mean-square for a $z$$\ge10$$\,$cm volume, where a large spread is
caused by the inefficient light collection from crystal back corners.
\end{enumerate}

For lateral surfaces painted with a $r_d$$\approx\,$0.9 diffuse substance 
without an air gap we find that the average photon collection
probability is lower, $\sim\,$17$\,$\%, and the axial light collection
nonuniformity as a function of axial position is always positive and is
usually larger than $+\,$10$\,$\% in a 22$\,$cm long detector, making
the root-mean-square of the 3D nonuniformity function $\ge\,$20$\,$\%.

The Monte Carlo results for the optically uniform half-hexagonal and 
trapezial CsI crystals with the same range of optical properties predict 
smaller light yields and somewhat higher light collection
nonuniformities:
\begin{enumerate}
\item The average photon collection probability with a two inch
photocathode is about 12$\,$\%.
\item The typical axial nonuniformity is positive in front, $+\,$5$\,$\%/10 
cm, and negative in the back crystal half, with a $-30\,$\%/10$\,$cm variation,
\item The simulated transverse light output from the front crystal half
is very similar to one for the full crystal shapes, increasing up to
$+\,$3$\,$\% away from the detector central axis, but decreasing by almost
a third in the back corners of the crystal.
\item The root-mean-square of detected light output varying between 
3.0$\,$\% and 4.5$\,$\%, in the front ($z$$\le$10$\,$cm) and back ($z$$\ge$10$\,$cm) 
crystal halves, respectively.
\end{enumerate}

These light collection probability distributions calculated for 
optically homogeneous crystals explain the major features of the 
measured optical nonuniformities presented below.

\section{Measured Nonuniformities of CsI Detector Light Responses}

The three-dimensional (3D) spatial distribution of the light output of a 
scintillation detector can
be specified by giving the number of photoelectrons $N_{\rm pe}(x,y,z)$
produced by 1$\,$MeV energy deposition at point ($x,y,z$) (``3D light 
nonuniformity function''). In the following discussion we limit 
ourselves to the linear one-dimensional variations of 
the detected light output separable in the axial and transverse directions:
\begin{eqnarray}
N_{\rm pe}(z,x)\propto\cases{
N_1+a_{z1}\cdot z+a_t(z)\cdot x,& $z\le10\ {\rm cm}$, and $x=\pm15\ 
{\rm cm}$ \cr
N_2+a_{z2}\cdot z+a_t(z)\cdot x,& $z\ge10\ {\rm cm}$, and $x=\pm15\ 
{\rm cm}$\cr
}
\end{eqnarray}
where $a_z$\/ ($a_t$) is the linear optical nonuniformity coefficient in
the $z$\/ ($x$) coordinate, and the coordinate system origin is at
the center of the detector front face, as explained in Sec.~\ref{tks}.

A simple and straightforward parameterization of the detector light
response nonuniformity can be made on the basis of scatter plots showing
the light output per unit pathlength as a function of longitudinal or
transverse cosmic muon-CsI crystal intersection coordinates. In this
analysis we take only cosmic muon events with almost perpendicular 
trajectories ($\theta_z$$\ge$85$^\circ$), so that the measured
two-dimensional light outputs are averaged over the pathlengths and
over $\sim\,1\,$cm$^2$ vertical cross sections. The cosmic muons
deposit the energy along the well-defined ionization track of
the length $d$\/ in the crystal. This description will somewhat
underestimate the real light collection probabilities (see Sec.~\ref{137s})
due to the integration of a three-dimensional light nonuniformity
function along the charged particle track. Our work on the fully 
3D reconstruction of the scintillator light response will be presented 
in a forthcoming publication~\cite{Frl96}.  

Fig.~\ref{fig:expected6} shows the axial variation of normalized ADC values
as a function of distance of energy deposition from the front crystal
face for the six representative crystals. The axial positions were
calculated by averaging two $z$\/ values of the cosmic muon track
intersection with the crystal surfaces. 

Panels on Figs.~\ref{fig:expected7a} and~\ref{fig:expected7c}
show the transverse dependence of measured light output per unit
pathlength, where the independent variable is the distance from
the crystal axis measured in the horizontal plane. The selected scatter
plots show the data points for the axial slices at $z_0$=6$\pm$1$\,$cm 
and $z_0$=18$\pm$1$\,$cm.

We have chosen to parameterize the axial light output nonuniformity with
two piecewise linear functions: the light output nonuniformity coefficient
$a_{z1}$($z$$\le$10$\,$cm) for the front half of the crystal, and 
the light nonuniformity coefficient $a_{z2}$($z$$\ge$10$\,$cm) for 
the back half of the crystal, both values expressed in \%/cm.
The transverse light output variation at a fixed 
axial distance $z_0$, $a_t(z_0)$, is described by the average value of 
a change in luminosity left and right of the crystal axis.

Several general features are readily noticeable: the spread of measured 
points due to energy deposition straggling, the gradual variation in
the collected light along and perpendicular to the detector axis, and
the decrease in detected light when the photon generation occurs
close to lateral detector surfaces and, in particular, near the back 
corners of the detector volumes.

The calculated ADC/pathlength data points shown on the panels of
Figs.~\ref{fig:expected6}--\ref{fig:expected7c} have been fitted with 
linear functions imposing the ``robust'' condition of the minimum
absolute deviation between the measured and calculated values, Eqs. 14.
The normalized light output scatter plots in the axial and transverse
coordinates, as well as two-dimensional ADC/$d$\/ distributions in the
$x$-$z$\/ bins and associated linear light output nonuniformity
coefficients have been documented for all studied CsI crystals and are
available for inspection at the PIBETA WWW site~\cite{pb}.
The average scintillation properties of all the measured CsI crystals are
listed in Tables~\ref{tab1}~and~\ref{tab2}.

In summary, for a set of fifty-nine full hexagonal and pentagonal
crystals we find:
\begin{enumerate}
\item The average axial light nonuniformities are
$a_{z1}$($z\le10$$\,$cm)=$-$0.1$\,$\%/cm 
and $a_{z2}$($z\ge10$$\,$cm)=$-$1.3$\,$\%/cm, respectively (see 
Fig.~\ref{fig:a1a2}).
\item Distribution of nonuniformity coefficients could be described 
by a Gaussian with $\sigma_{a_{z1}}$$\approx\,$1.3$\,$\%/cm.
\item The average transverse light output is flat for $z$$\le$10$\,$cm,
and the typical light variation is $-15\,$\% at $z$=18$\,$cm.
\item The Kharkov-grown crystals on average have twice the axial optical 
nonuniformity of the Bicron-grown crystals.
\end{enumerate}

For fifteen half-hexagonal and trapezial crystal shapes we find:
\begin{enumerate}
\item The average axial nonuniformities are $a_{z1}$($z$$\le$10 
cm)=$-\,$0.3$\,$\%/cm and $a_{z2}$($z$$\ge$10$\,$cm)=$-$1.0$\,$\%/cm.
\item The transverse light output variation is limited in the front 
crystal half, but increases up to $-\,$30$\,$\% in the back of the crystal.
\end{enumerate} 

With the insights gained in the simulation of the light collection
probabilities (Sec.~\ref{tks}) we conclude that the $\pm\,$2$\,$\%/cm spread 
of the nonuniformity coefficients corresponds to an equivalent range of
0.8--1.0 in crystal surface reflectivities, and/or 100--250$\,$cm range 
in CsI attenuation lengths.

\section{Light output uniformity scans of CsI crystals with 
$^{137}$Cs gamma source}\label{137s}

A light-tight plywood box was made to house a single CsI detector and
associated measurement apparatus described below. A 0.662$\,$MeV $^{137}$Cs 
gamma source was placed on a moveable support next to the detector and 
collimated by a 20$\times$10$\times$5$\,$cm$^3$ lead brick with a 6$\,$mm 
collimator hole.
The photomultiplier analog signal from the detector was processed with
an ORTEC 454 timing amplifier with a gain of 30 and a 50 ns
integration time constant. The amplifier output was digitized with
a peak sensing ORTEC AD811 ADC unit. The same signal was discriminated
and produced the trigger rate of about 5 kHz with a $^{137}$Cs source
present. The background rate, without the source present, was 50--100 Hz.

Pedestal runs were taken with a clock trigger and used to properly
correct offsets of the ADC spectra. Temperature variation during one run
was less than 0.3$^\circ$C, typically causing the overall light output 
variation of less than 0.5$\,$\% in a ADC spectrum that was gated with a 
100 ns window. Several runs were taken with the source removed to find the 
shape of the background spectrum. It was confirmed that the background 
spectrum does not depend on the position of the lead collimator, so 
the same background lineshape was used in the analysis for all source 
positions. 

The collimated $^{137}$Cs source was placed by remote control in turn at 
five points along the axis of each crystal at 2, 6, 10, 15 and 20$\,$cm 
from the front face of the crystal.

Following background subtraction, recorded spectra were fitted with
a sum of a Gaussian and exponential function, Fig.~\ref{fig:cs137}.
The peak position and the FWHM for each spectrum were extracted with the 
statistical uncertainty lower than 0.3$\,$\%.

The dependence of the peak position on the placement of the source
is illustrated in Fig.~\ref{fig:cs137f} and follows the trend of the
tomography results. The {\tt GEANT} simulations revealed that the 662
keV gamma rays could probe all the volume of CsI crystal, but
the shower energy deposited at the central axis is only about one-sixth
of the energy deposition near the crystal surface.
The cosmic muons transfer the energy uniformly along their tracks in CsI
material: the scintillation volume over which the measured light output
is integrated is therefore larger, and inherent averaging leads to
smaller extracted optical nonuniformity coefficients.
This feature is borne out in the panels of Fig.~\ref{fig:cs137} where the
tomography data for three different crystals yield consistently smaller
axial light nonuniformities when compared with the radioactive source
measurements. 

The simple $^{137}$Cs scans with the described apparatus were used 
to evaluate the light collection nonuniformities for up to six crystals 
per day.

Radioactive source scan measurements similar to ours are described 
in Refs.~\cite{Bro95,Dow90}. 

\section{A {\tt GEANT} simulation of the PIBETA calorimeter response to
10--120$\,$MeV positrons and photons}

We have studied the simulated PIBETA calorimeter response to the 69.8$\,$MeV
$\pi^+$$\rightarrow$$e^+\nu$\/ positrons and to the (69.8+2$m_{e^+}$) 
MeV photons, where $m_{e^+}$\/ is the positron rest mass. 
The goal was to find the intrinsic difference between the 
responses to monoenergetic positrons and photons with identical 
total incident energies, emphasizing the correct modeling of
the low-energy tail below the edge of the Michel
$\mu$$\rightarrow$$e^+\nu\nu$\/ spectrum at 52.8$\,$MeV.

The 0.46$\,$\% accuracy of the new TRIUMF measurement~\cite{Bri94} of the 
$\pi^+$$\rightarrow$$e^+\nu$\/ branching ratio was limited 
by event statistics and systematic uncertainties in evaluating 
the low-energy tail of the positron peak. 
The $\pi^+$$\rightarrow$$e^+\nu\gamma$\/ positrons and the forward-peaked 
bremsstrahlung gammas were detected in a 46-cm-diameter$\times$51-cm-long
NaI(Tl) crystal ``TINA''. 
The tail correction of 1.44$\pm\,$0.24$\,$\% for the energy region from 
0$\,$MeV to 52.3$\,$MeV was determined by subtracting the measured Michel
positron spectrum from the detector positron response functions measured 
with the 20--85$\,$MeV $e^+$\/ monoenergetic beams. An additional tail 
component of $\sim\,$0.4$\,$\% due to radiative processes was estimated by 
Monte Carlo method. The TRIUMF group apparently made no attempt to 
account for the potential difference in the scintillator light response 
of positrons and photons and neglected potential nonlinearities of
the energy scale. Both effects could arise from the light collection 
nonuniformities of the NaI(Tl) detector.

In the recent PSI  $\pi^+$$\rightarrow$$e^+\nu$\/ experiment~\cite{Cza93} 
the quoted 0.29$\,$\% systematic uncertainty of the extracted branching ratio
was also dominated by the 1.64$\pm\,$0.09$\,$\% electromagnetic loss
fraction as well as by the 0.95$\pm\,$0.19$\,$\% 
contribution from the photonuclear reactions. The $\sim\,$4$\pi$
calorimeter used in that measurement was made of 132 hexagonally shaped
BGO crystals, with the light yield claimed to be homogeneous to within
1.5$\,$\% over the whole 20$\,$cm length of every crystal.

The lineshapes of positrons and photons in our PIBETA calorimeter and
their dependence on the optical properties of 240 constituent CsI
crystals were predicted using the {\tt GEANT} detector description and
simulation tools. The code defined the geometry and tracking media for
the complete PIBETA detector~\cite{Pib95}. The inner detector region
was occupied by two cylindrical wire chambers and a segmented plastic
veto detector. The incident particles, monoenergetic positrons and
photons, were generated in the center of the crystal sphere.

Photonuclear reactions in the active detectors as well as in
the surrounding passive material were modeled in a user
subroutine~\cite{Frl95} that was added to the default {\tt GEANT}
version {\tt 3.21}. The probabilities of photonuclear reactions were
calculated using the published cross sections from the $\gamma N$
reaction thresholds up to the energy of 120 
MeV~\cite{Ahr75,Lep81,Heb76,Ber69,Bra66,Jon68}.  

The 240 individual CsI module shapes were specified, taking into account 
the mechanical tolerances of the physical CsI crystals, with Teflon and
aluminized Mylar wrappings filling the 200 $\mu$m intermodule gaps. 
The irregular CsI modules had to be constructed from up to six
{\tt GEANT} generalized trapezoidal wedges. Every CsI crystal volume 
was considered a sensitive detector with the associated luminosity
$\bar N_{\rm pe}$\/ expressed as the number of photoelectrons per MeV, and two
axial and one transverse light collection nonuniformity coefficients,
$a_{z1}$, $a_{z2}$, and $a_t$, respectively. A set of seventy-four
detector luminosity and nonuniformity values, extracted in
the tomography analysis, were initialized in the {\tt GEANT}-accessible
database. The optical parameters of the remaining 163 modules were drawn
randomly from the distributions in Figs.~\ref{fig:npf} 
and~\ref{fig:a1a2}. The assumption was that the crystals produced in 
the future will be of the same optical quality as the ones
that are already delivered.

We have always simulated at least 10$^5$ events for every fixed 
set of the calorimeter parameters 
($\bar N_{\rm pe}$,$a_{z1}$,$a_{z2}$,$a_t$). The low-threshold trigger was 
defined by requiring a sum of the ADC readings
exceed 5$\,$MeV of the light-equivalent energy for one calorimeter
supercluster containing 32 crystals. The integrated detector acceptances
with a low-level trigger were 85.8$\,$\% for $\sim\,$70$\,$MeV positrons 
and 85.5$\,$\% for $\sim\,$70$\,$MeV photons, so all extracted quantities had
the relative statistical uncertainties of $\sim\,$0.2$\,$\%. 
The response of the calorimeter was parameterized by the FWHM 
of the simulated ADC spectrum and the tail contribution being between
the adjustable low level (default {\tt LT} being 5$\,$MeV) and high level
thresholds (two default values {\tt HT$_1$} and {\tt HT$_2$} being 54$\,$MeV 
and 55$\,$MeV, respectively).

For the purposes of comparison and gain normalization we first studied
the response of an ideal, homogeneous calorimeter, with
realistic CsI crystal light outputs. The ``software'' gain factor of
every individual CsI detector was determined from the fitted peak
positions of the simulated ADC spectrum sum over the crystal with
maximum energy deposition and its nearest neighbors. The peak value
positions were found after smoothing the histograms with a multiquadric
function, to improve on the limited simulation statistics. The ratios
of peak positions in the ideal, homogeneous calorimeter and
the corresponding values for the physical, nonuniform calorimeter, were,
by definition, the individual detector gain corrections. The values of
these ratios were refined in three steps of the iterative procedure.
The extracted energy resolution of the nonuniform detector in principle
depends on the convergence of the gain-matching process and
the resulting uncertainties of the individual detector gains.
That gain-matching process was equivalent to the PMT high voltage
adjustments in the real experiment that are effected to obtain
the best energy resolution with the modular detector. We estimate
that our simulation procedure fixes the software detector gains with
the accuracy of $\sim\,$1$\,$\%.

The set of gain constants depended on the optical properties of our
CsI crystals as defined in our {\tt GEANT} database, but also upon the  
incident particle chosen for the Monte Carlo calibration runs and its 
energy because of differences in shower developments of photons and positrons. 
The gain normalizations of the calorimeter detectors were calculated  by
aligning the simulated $\pi^+$$\rightarrow$$e^+\nu$\/ positron peak positions. 
The same procedure for matching the CsI detector gains was used in the
real data-taking runs. Due to the gain
renormalization, the influence of transverse light nonuniformities
smaller than 10$\,$\% on the ADC spectrum peak positions and the ADC
lineshape could be neglected in comparison with the axial light
nonuniformity effects. The simulated ADC values were summed over all
detectors with the energy deposition above the 1$\,$MeV TDC threshold.
We also examined the calculated ADC sums for the clusters which contained
the crystal with the maximum energy deposition and its nearest neighbors.
The full-widths at half maximum for these spectra are labeled in
the following figures and tables as FWHM$_{(220)}$ and FWHM$_{\rm (NN)}$,
respectively. No event-to-event uncertainties of the ADC pedestal values
were assumed in the simulation. Stability of the PIBETA electronics tested
under real experimental conditions in 1996 calibration runs and
the quality of the algorithms used for the first and second pedestal
correction reduces the pedestal peak root-mean-square to
$\sim\,$8 channels, equivalent to $\sim\,$0.3$\,$MeV.

The {\tt GEANT}-calculated resolution of the PIBETA calorimeter
consisting of 220 CsI detector crystals and 20 veto crystals was
parameterized by the fractional full width at half maximum: 

\begin{eqnarray}
{ {\rm FWHM_{(220)}}\over {E_{e^+}} }(\%)&=&  
{ {2.36\cdot\sigma_{E_{e^+}} }\over {E_{e^+}} }= \\
&=&\cases{ {{(54.69+1.197 E_{e^+}+0.4656\cdot 10^{-2}E_{e^+}^2)}
/ {E_{e^+}^{0.8030}} }\cr
{ {(37.19+0.060 E_{e^+}+0.2734\cdot 10^{-2}E_{e^+}^2)}
/ {E_{e^+}^{0.5256}} }\cr} \
\end{eqnarray}
for the positrons with the most probable peak energy $E_{e^+}$\/ MeV and
\begin{eqnarray}
{ {\rm FWHM_{(220)}}\over {E_\gamma} }(\%)=\cases{
{ {(37.07+0.5507 E_\gamma+0.3233\cdot 10^{-2}E_\gamma^2)}
/ { E_\gamma^{0.6549}} }\cr
{ {(40.52+0.9437\cdot E_\gamma+0.4931\cdot 10^{-2}E_\gamma^2)}
/ { E_\gamma^{0.6942}} }\cr}
\end{eqnarray}
for the monoenergetic photons with the detected peak energy $E_\gamma$\/ 
MeV (Fig.~\ref{fig:fit_fwhm_220}).
The top lines refer to the cases of an optically uniform detector ($a_z$=0)
while the bottom equations describe the predicted 
response of the nonuniform calorimeter ($a_z$={\tt TOMOGRAPHY}).

The average fractional energy resolution FWHM$_{(220)}$ of 5.3$\,$\% 
(6.0$\,$\%) 
was achieved for the $\sim\,$70$\,$MeV kinetic energy positrons in 
the optically uniform (nonuniform) calorimeter, as compared to 5.7$\,$\%
(6.8$\,$\%) resolution for the equivalent energy photons. That was
the resolution found with no cuts applied on the light-equivalent energy
generated in the calorimeter vetoes. Requiring less than 5$\,$MeV detected
in the veto shield decreased the statistics by about 10$\,$\% and
suppressed the low-energy tail, but did not improve the energy resolution
at the peak position.

Limiting the ADC sums to the nearest-neighbor crystals (six or seven
crystal clusters) the fractional FWHM$_{\rm (NN)}$\/ energy resolution 
could be parameterized by:
\begin{eqnarray}
{ {\rm FWHM_{\rm (NN)}}\over {E_{e^+}} }(\%)=
\cases{ {{(69.19+3.057 E_{e^+}+0.4198\cdot 10^{-2}E_{e^+}^2)}
/ {E_{e^+}^{0.8560}} }\cr
{ {(68.13+2.479 E_{e^+}+0.3368\cdot 10^{-2}E_{e^+}^2)}
/ {E_{e^+}^{0.8018}} }\cr} 
\end{eqnarray}
for the positrons with the detected peak energy $E_{e^+}$\/ MeV and
\begin{eqnarray}
{ {\rm FWHM_{\rm (NN)}}\over {E_\gamma} }(\%)=\cases{
{ {(43.16+2.318 E_\gamma+0.4148\cdot 10^{-2}E_\gamma^2)}
/ { E_\gamma^{0.7723}} }\cr
{ {(39.57+2.302 E_\gamma+0.1329\cdot 10^{-2}E_\gamma^2)}
/ { E_\gamma^{0.7199}} }\cr}
\end{eqnarray}
for the monoenergetic photons with the detected peak energy $E_\gamma$\/ 
MeV (Fig.~\ref{fig:fit_fwhm_nn}).

The average FWHM$_{\rm (NN)}$\/ for the 70$\,$MeV positrons and gammas in 
nearest-neighbor nonuniform crystal clusters was 8.7$\,$\% (6.1$\,$MeV) and
9.9$\,$\% (6.8$\,$MeV), respectively. Imposing the 5$\,$MeV cut on the detector
veto signals improves these resolutions only marginally. These numbers
should be compared with the response width of the ideal uniform
nearest-neighbor clusters: 8.2$\,$\% (5.7$\,$MeV) and 8.7$\,$\% (6.2$\,$MeV).

The percentage of the events in the tail between the preset low and high 
energy threshold was tracked in the same {\tt GEANT} simulation. 

The listing of the low-energy tail contributions for 69.8$\,$MeV $e^+$\/ and
70.8$\,$MeV $\gamma$\/ in Table~\ref{tab3} shows that in the optically 
uniform calorimeter they differ by 1.8$\,$\%. The optical nonuniformity does 
not change the positron tail contribution, but increases the photon tail
by $\sim\,$0.2$\,$\%. The imposition of the 5$\,$MeV veto-shield cut decreases
the electromagnetic leakage to 2.4$\,$\% for positrons and 4.7$\,$\% for 
photons, a $e^+$-$\gamma$\/ tail difference of $+\,$2.3$\,$\%.

Table~\ref{tab4} shows the low-energy tail components for the 
simulated nearest-neighbor cluster ADC sums: with no applied cuts the
contributions were 4.7$\,$\% and 6.7$\,$\% for $e^+$'s and $\gamma$'s,
respectively. The veto shielding cuts decreased both fractions by about
0.3$\,$\%.

If the positron originates from a radiative
$\pi^+$$\rightarrow$$e^+\nu\gamma$\/ decay, its average Monte Carlo peak
position is essentially unchanged. Its tail contribution is unaffected
if simulated ADC sums extend over full calorimeter but are increased by
$\sim\,$1.4$\,$\% if only the nearest-neighbor clusters are summed. 
The radiative decay matrix elements used in the calculation were taken
from Ref.~\cite{Bro64}. All quoted numbers have
an approximate systematic uncertainty of $\sim\,$0.2$\,$\%.

The detector response to positrons and photons in the energy range
between 10$\,$MeV and 120$\,$MeV is also nonlinear because the volume
distribution of the shower energy deposition depends on the incident
particle type and the energy. Our {\tt GEANT} studies showed that
the magnitude and energy dependence of these nonlinearities does not
change significantly because of CsI crystal light collection nonuniformity.
Results of the calculations, both for the case of a homogeneous and
optically nonhomogeneous calorimeter, are displayed in 
Figs.~\ref{fig:fit_nonlin_220} and~\ref{fig:fit_nonlin_nn}.
Nonlinearities of the detector response in the covered energy range for
positrons and photons are very close in magnitude and shape of
energy variation and amounted to $\sim\,$1.5$\,$\% for the ADC sums over
220 crystals and up to $\sim\,$2.8$\,$\% if the simulated ADC sums were
restricted to the over-the-threshold ADCs of nearest neighbors. 

\section{Conclusion}

We have measured the optical properties of seventy-four pure cesium iodide 
crystals that were polished and wrapped in the diffuse Teflon reflector. 
The results are summarized separately for the full-sized and half-sized
crystals in Tables~\ref{tab1} and~\ref{tab2}. 

The deduced light yields parameterized by two axial and one transverse light 
collection nonuniformity coefficients constitute a minimum set of 
parameters necessary for a realistic Monte Carlo simulation of 
the modular CsI calorimeter.

The predicted energy resolutions FWHM$_{(220)}$ for $\sim\,$70$\,$MeV
positrons and photons in the full PIBETA calorimeter with the ideal,
optically uniform CsI modules with specified luminosities were shown
to be close, 3.7$\,$MeV and 4.0$\,$MeV, respectively. The upper limit of
the low energy tail contributions in the region between 5$\,$MeV and 55$\,$MeV
were calculated to be 6.9$\,$\% and 8.7$\,$\% for the positrons and gammas in
an optically homogeneous detector, respectively. After applying
the 5$\,$MeV calorimeter veto cut, these tail corrections decrease
to 2.2$\,$\% and 4.5$\,$\%.

The average deduced axial light nonuniformities of real CsI crystals
wrapped in a Teflon sheet had negative slopes, $-\,$0.18$\,$\%/cm and
$-\,$1.6$\,$\%/cm, for the front and back half of the crystal volume,
respectively. The corresponding {\tt GEANT} simulation of the nonuniform
PIBETA apparatus for 68.9$\,$MeV $e^+$\/ and 70.8$\,$MeV $\gamma$'s shows that
both energy responses will be broadened to the average
FWHM$_{(220)}$ of 5.9$\,$\% and 6.9$\,$\% while the associated tail
contributions will change only for photon spectra, increasing
the low-energy tail by $\sim\,$0.3$\,$\%. The simulated calorimeter
ADC spectra are shown in Figs.~\ref{fig:tails1} and~\ref{fig:tails2}.

The nonlinearity of the measured energy scale caused by the optical
nonuniformity is $\le\,$2.8$\,$\% throughout the relevant $e^+$/$\gamma$
energy range. This spread is consistent with the precision of energy 
calibration required to extract
the tail corrections with the systematic uncertainty of $\sim\,$0.2$\,$\%. 

The predicted ADC spectra of the monoenergetic positrons, electrons
and tagged photons in the energy range 10--70$\,$MeV will be compared with
the measured responses of the partial CsI calorimeter arrays in a 
forthcoming publication~\cite{Frl97}.

\section{Acknowledgements}
The authors wish to thank Micheal Sadler of the Abilene Christian
University (ACU) for lending us the drift chamber tomography apparatus.
Derek Wise, also of ACU, has helped with the cosmic muon tomography
measurements.  Roger Schnyder of Paul Scherrer Institut has maintained and
repaired the faulty electronics modules.  Their help is gratefully
acknowledged.

This work is supported and made possible by grants from the US National
Science Foundation and the Paul Scherrer Institute.
\bigskip

\vfill\eject

\clearpage


\begin{figure}[!tpb]
\caption{The PIBETA CsI shower calorimeter consists of
240 tapered hexagonal, pentagonal and trapezial pyramids.
The crystal ball has an opening for beam entry and a symmetric one
for light readout from inner detectors (not shown).} 
\label{fig:ball}
\end{figure}

\begin{figure}[!tpb]
\caption{The dimensioned drawing of the HEX--A (top) and HEX--D1 (bottom)
CsI crystal shapes.}
\label{fig:shape}
\end{figure}
 
\begin{figure}[!tpb]
\caption{The fast-to-total light component ratio for the all studied CsI 
crystals. The light outputs from optically polished and unwrapped 
crystals were measured immediately after the crystals were received 
from a manufacturer with a digital oscilloscope at the average 
ambient temperature of 22$^\circ$C.} 
\label{fig:ft}
\end{figure}
 
\begin{figure}[!tpb]
\caption{The Monte Carlo simulation of penetrating cosmic muons 
interacting with the experimental apparatus. Six CsI crystals were
defined inside a dark box and deposited energies and track lengths
of cosmic muons and generated secondary particles in detectors were 
digitized.}
\label{fig:drift}
\end{figure}
 
\begin{figure}[!tpb]
\caption{The schematic diagram of the electronics logic for the 
tomography apparatus. The calibration event was triggered by two-fold 
scintillator coincidence S1$\times$S2. The trigger used during cosmic
muon data acquisition required three-fold coincidence between two 
scintillators and at least one CsI detector, (S1$\times$S2)$\times$$\sum 
{\rm CsI}_i$.}
\label{fig:tmelec1}
\end{figure}
 
\begin{figure}[!tpb]
\caption{One-dimensional residuals defined as differences between the 
best straight line fits from the reconstructed coordinates in the hit
chambers 1 and 2 and the measured positions in the top chamber 3. 
The FWHM of the peak is 0.9 mm for $x$\/ coordinate and 1.5
mm for $y$\/ coordinate, respectively. The 1.2$\,$cm wide plateaus contains
$\sim\,$5$\,$\% events with incorrect left-right identification
and/or the events undergoing the large angle scattering in the apparatus.
}
\label{fig:deviation}
\end{figure}
 
\begin{figure}[!tpb]
\caption{The Monte Carlo distribution of scattering angles of the
cosmic muons intersecting a HEX-A CsI crystal shape. The root-mean-square
of the scattering angle for the events with the reconstructed tracks is
0.66$^\circ$. That value corresponds to the CsI pathlength uncertainty of
$\sim\,$1$\,$mm.}
\label{fig:ths}
\end{figure}

\begin{figure}[!tpb]
\caption{The average measured cosmic muon pathlength in CsI crystals
was $\sim\,$6$\,$cm. An accurate determination of the track length requires
the precise knowledge of the absolute crystal position with respect to
the drift chambers. The number of muon tracks intersecting
a CsI crystal volume is shown as a function of horizontal crystal 
offsets. Contours of the 25-level plot reveal a crystal position with
an uncertainty of $\pm\,0.05\,$cm (left panel, pentagonal CsI crystal: 
$x_{\rm off}=12.32$$\pm\,$$0.05\,$cm, $z_{\rm off}=30.50$$\pm\,$$0.05\,$cm).
The same type of plot is shown for one hexagonal CsI crystal on 
the right panel.}
\label{fig:position22}
\end{figure}
\clearpage
 
\begin{figure}[!tpb]
\caption{Fast light output temperature coefficients for all analyzed
CsI calorimeter detectors. The mean value is $-\,$1.4$\,$\%/$^\circ$C,
with the temperature coefficients ranging from $-\,$4$\,$\%/$^\circ$C to 
$+\,$4$\,$\%/$^\circ$C. Temperature changes inside the light-tight box (filled 
markers) and average ADC values (open circles) for one detector with the
negative temperature coefficient of $-$1.9/$^\circ$C are shown in 
the inset. Data points shown in the inset span one week of data 
acquisition.}
\label{fig:tf}
\end{figure}

\begin{figure}[!tpb]
\caption{LED spectra in one CsI detector recorded for five 
different values of the LED driving voltage. The horizontal scale is
a raw ADC count of the fast CsI scintillation component (a 100 ns ADC gate)
corrected for the pedestal offset. The equivalent energy range covered is
10--100$\,$MeV. The photoelectron statistics can be determined by measuring
the width of a photodiode peak $\sigma^2_E$\/ as a function of 
a spectrum mean value $\bar E$. The linear dependence described by Eq.
(\ref{eq:led}) is demonstrated in the inset. The inverse slope of 
the straight line equals the number of photoelectrons created per unit 
energy deposition.}
\label{fig:set2}
\end{figure}
 
\begin{figure}[!tpb]
\caption{The number of photoelectrons per MeV of deposited
energy for all studied CsI calorimeter modules. The left panel shows
the distribution for the full-sized hexagonal and pentagonal CsI 
detectors viewed with three inch phototubes. The right panel is
the equivalent histogram for the smaller half-hexagonal and trapezial
detectors equipped with two inch phototubes.}
\label{fig:npf}
\end{figure}
 
\begin{figure}[!tpb]
\caption{The simulated light collection probability as a function of
axial ($z$) and transverse ($x$) coordinates for an optically ideal hexagonal
detector (HEX--A) with a two-layer Teflon wrapping. The size of the
histogrammed two-dimensional bin is 1$\times$1$\,$cm$^2$. The average light 
collection probability with a three inch phototube is 23$\,$\%.}
\label{fig:hexa_9}
\end{figure}
 
\begin{figure}[!tpb]
\caption{The Monte Carlo light collection probability as a two 
dimensional function of axial and transverse coordinates for an ideal
half-hexagonal detector (HEX--D1) with a Teflon wrapping. The average
light collection probability with a two inch photocathode is 12$\,$\%.}
\label{fig:hh1d_9}
\end{figure}
 
\begin{figure}[!tpb]
\caption{Relative light per unit pathlength ADC/$d$\/ as a function of
axial position of energy deposition is shown for six CsI crystals.
The phototubes were always positioned on the surface with
the larger values of the $z$\/ coordinate. Plotted values were
corrected for temperature variation effects and the PMT gain drifts.} 
\label{fig:expected6}
\end{figure}
 
\begin{figure}[!tpb]
\caption{Detected light per unit pathlength as a function of
transverse coordinates of energy deposition shown at the fixed distance 
along the crystal axis ($z_0$=6$\pm\,1\,$cm) for six CsI detectors.}
\label{fig:expected7a}
\end{figure}

\begin{figure}[!tpb]
\caption{Detected light per unit pathlength as a function of transverse
position of energy deposition shown at the fixed axial distance
($z_0$=18$\pm1\,$cm) for six CsI crystals.}
\label{fig:expected7c}
\end{figure}
\clearpage 

\begin{figure}[!tpb]
\caption{The light nonuniformity coefficients $a_{z1}$\/ and $a_{z2}$
(\%/cm), determined separately for front ($z\le$10$\,$cm) and back
($z\ge$10$\,$cm) crystal sections, are shown here in a scatter-plot for
74 CsI detectors. The full-sized hexagonal and pentagonal detectors are
represented with the full marker points, while the open circles indicate
the half-hexagonal and trapezial detector shapes. The dotted line is
the result of a ``robust'' straight-line fit:
$a_{z1}$(\%/cm, $z$$\le\,$10$\,$cm)=$-$1.44+1.19$\cdot a_{z2}$(\%/cm,
$z$$\ge$10
cm).}
\label{fig:a1a2}
\end{figure}

\begin{figure}[!tpb]
\caption{The 662 keV $^{137}$Cs gamma source and the background spectrum
measured at the source axial position $z$=10$\,$cm from the front face
of the pentagonal S003 crystal.}
\label{fig:cs137}
\end{figure}

\begin{figure}[!tpb]
\caption{The comparison of normalized light output (ADC 
reading/pathlength) along the axis of three different CsI crystals
deduced from the cosmic muon tomography data (full line), and as
measured with the $^{137}$Cs gamma source scan (dotted line).}
\label{fig:cs137f}
\end{figure}
 
\begin{figure}[!tpb]
\caption{The predicted fractional energy resolution ${\rm FWHM_{(220)}}/
E_{(220)}$ of the PIBETA calorimeter for positrons and photons in
the energy range 10--120$\,$MeV. The {\tt GEANT}-deduced relative
FWHM$_{(220)}$ for the sum over 220 CsI ADCs is shown both for
an ideal homogeneous detector (full lines) as well as for
a nonuniform calorimeter (dotted lines). Notice that the zero value on 
the vertical scale is suppressed. The curves represent 
the parameterizations quoted in the text.}
\label{fig:fit_fwhm_220}
\end{figure}

\begin{figure}[!tpb]
\caption{The {\tt GEANT}-deduced ${\rm FWHM_{\rm (NN)}}/E_{\rm (NN)}$
for the sum over nearest-neighbor CsI ADCs is shown both for an ideal
homogeneous detector as well as for a nonuniform calorimeter.}
\label{fig:fit_fwhm_nn}
\end{figure}
 
\begin{figure}[!tpb]
\caption{Nonlinearity of the PIBETA calorimeter energy response for
positrons and photons in the energy range 10--120$\,$MeV.
The ratio of the detected ADC peak position to the incident particle
energy is calculated in a {\tt GEANT} simulation. Results for
an ideal homogeneous detector and a nonuniform calorimeter are shown
separately. The normalization is given by ADC/$E$(70$\,$MeV $e^+,\gamma$)=1.}
\label{fig:fit_nonlin_220}
\end{figure}
 
\begin{figure}[!tpb]
\caption{Nonlinearity of the PIBETA calorimeter energy response for
positrons and photons in the energy range 10--120$\,$MeV. Ratio of
the detected peak position of summed nearest-neighbor ADC spectra
to the incident particle energy is calculated in a {\tt GEANT}
simulation.}
\label{fig:fit_nonlin_nn}
\end{figure}
\clearpage 

\begin{figure}[!tpb]
\caption{The predicted PIBETA calorimeter spectra of the monoenergetic
69.8$\,$MeV positrons and 70.8$\,$MeV photons. The 220 ADC values above the
1.0$\,$MeV TDC threshold were summed. A {\tt GEANT} simulation assumed
different numbers of photoelectrons/MeV and linear axial light
collection nonuniformities as extracted in the tomography analysis.
The results for the ``realistic'' detector depend on the treatment of
the CsI crystal surfaces and the details of the gain matching algorithm.}
\label{fig:tails1}
\end{figure}

\begin{figure}[!tpb]
\caption{The predicted PIBETA calorimeter spectra of the monoenergetic
69.8$\,$MeV positrons and 70.8$\,$MeV photons. The ADC value for the CsI crystal
with the maximum ADC value and the ADC readings of its nearest
neighbors were summed. A {\tt GEANT} simulation assumed different
numbers of photoelectrons/MeV and linear axial light collection
nonuniformities for each detector, as measured in the tomography apparatus.}
\label{fig:tails2}
\end{figure}
\vspace*{\stretch{2}}\clearpage

\bigskip
\begin{table}[!pbt]
\caption{Average scintillation properties of hexagonal and pentagonal
PIBETA CsI calorimeter shapes PENTAs, HEX--As, HEX--Bs, HEX--Cs
and HEX--Ds (59 crystals). All crystals were polished and wrapped in two
layers of a Teflon membrane and one layer of aluminized Mylar film.
The light yields are normalized to the temperature of 18$^\circ$C. All 
other parameters were measured at the average laboratory room
temperature of 22$^\circ$C.}

\label{tab1}
\begin{tabular}{lcc}
\hline
\multicolumn{1}{c}{\ } 
&\multicolumn{1}{c}{Bicron CsI} 
&\multicolumn{1}{c}{Kharkov CsI}\\
\multicolumn{1}{c}{\ } 
&\multicolumn{1}{c}{Crystals (22)} 
&\multicolumn{1}{c}{Crystals (37)}\\
\hline\hline
Fast-to-Total Ratio (100 ns/1 $\mu$s gate)   &  0.835    & 0.739   \\
\# Photoelectrons/MeV (100 ns ADC gate)      &  83.4     & 65.1    \\
\# Photoelectrons/MeV (1 $\mu$s ADC gate)    &  99.9     & 88.0    \\
Fast Light Temp. Coefficient (\%/$^\circ$C)  &  $-1.20$  & $-1.61$ \\
Total Light Temp. Coefficient (\%/$^\circ$C) &  $-1.25$  & $-1.40$ \\
Axial Nonuniformity Coefficient (\%/cm),     &           &         \\
\ \ \ $z$$\le$10$\,$cm, 100 ns ADC gate         &  $-0.200$ & $-0.124$\\
Axial Nonuniformity Coefficient(\%/cm),      &           &         \\
\ \ \ $z$$\ge$10$\,$cm, 100 ns ADC gate         &  $-1.62$  & $-1.61$ \\
Transverse Nonuniformity Coefficient(\%/cm), &           &         \\
\ \ \ $z_0$$=$6$\,$cm, 100 ns ADC gate            &  $-0.400$ & $-0.500$\\
Transverse Nonuniformity Coefficient(\%/cm), &           &         \\
\ \ \ $z_0$$=$18$\,$cm, 100 ns ADC gate           &  $-1.10$  & $-1.20$ \\
\hline
\end{tabular}
\end{table}
\clearpage

\bigskip
\begin{table}[!tbp]
\caption{Average scintillation properties of half-hexagonal and trapezial
PIBETA CsI calorimeter shapes HEX--D1/2s and VETO--1/2s (15 crystals).
All crystals were polished and wrapped in two layers of Teflon foil and
one layer of aluminized Mylar.}
\label{tab2}

\begin{tabular}{lcc}
\hline
\multicolumn{1}{c}{\ } 
&\multicolumn{1}{c}{Bicron CsI} 
&\multicolumn{1}{c}{Kharkov CsI}\\
\multicolumn{1}{c}{\ } 
&\multicolumn{1}{c}{Crystals (3)} 
&\multicolumn{1}{c}{Crystals (12)}\\
\hline\hline
Fast-to-Total Ratio (100 ns/1 $\mu$s gate)   &  0.806    & 0.728   \\
\# Photoelectrons/MeV (100 ns ADC gate)      &  34.5     & 31.5    \\
\# Photoelectrons/MeV (1 $\mu$s ADC gate)    &  42.8     & 43.3    \\
Fast Light Temp. Coefficient (\%/$^\circ$C)  &  $-2.08$  & $-0.63$ \\
Total Light Temp. Coefficient (\%/$^\circ$C) &  $-1.78$  & $-0.69$ \\
Axial Nonuniformity Coefficient (\%/cm),     &           &         \\
\ \ \ $z$$\le$10$\,$cm, 100 ns ADC gate         &  $-0.326$ & $-0.500$\\
Axial Nonuniformity Coefficient(\%/cm),      &           &         \\
\ \ \ $z$$\ge$10$\,$cm, 100 ns ADC gate         &  $-1.00$  & $-1.83$ \\
Transverse Nonuniformity Coefficient(\%/cm), &           &         \\
\ \ \ $z_0$$=$6$\,$cm, 100 ns ADC gate            &  $-0.755$ & $-0.827$\\
Transverse Nonuniformity Coefficient(\%/cm), &           &         \\
\ \ \ $z_0$$=$18$\,$cm, 100 ns ADC gate           &  $-4.10$  & $-5.14$ \\
\hline
\end{tabular}
\end{table}\clearpage

\bigskip
\begin{table}[!tbp]
\caption{The predicted energy resolutions and tail contributions for
69.8$\,$MeV $e^+$\/ and 70.8$\,$MeV $\gamma$'s events in the full PIBETA
calorimeter. The light output in photoelectrons/MeV and the linear axial
light collection nonuniformities measured for individual CsI crystals 
($a_{z1,z2}$={\tt TOMOGRAPHY}) were used in a {\tt GEANT} simulation. 
The values for the perfect optically homogeneous crystals 
($a_{z1,z2}$=0) were shown for comparison.}

\label{tab3}
\begin{tabular}{lcccc}
\hline
\multicolumn{1}{c}{Parameter} 
&\multicolumn{1}{c}{69.8$\,$MeV $e^+$}
& 
&\multicolumn{1}{c}{70.8$\,$MeV $\gamma$} 
& \\ 

\multicolumn{1}{c}{No $E_V$\/ cut} 
&\multicolumn{1}{c}{$a_{z1,z2}$=0} 
&\multicolumn{1}{c}{{\tt TOMOGRAPHY} $a_{z1,z2}$} 
&\multicolumn{1}{c}{$a_{z1,z2}$=0} 
&\multicolumn{1}{c}{{\tt TOMOGRAPHY} $a_{z1,z2}$} \\
\hline\hline
Peak Position (MeV)         & 68.87$\pm$0.03& 68.90$\pm$0.03
 & 70.10$\pm$0.03& 69.97$\pm$0.03 \\
FWHM$_{(220)}$ (MeV)        &  3.66$\pm$0.03&  4.10$\pm$0.03
 &  3.98$\pm$0.03&  4.76$\pm$0.03 \\
5$\le$Events$\le$54$\,$MeV (\%)&  6.44$\pm$0.09&  6.46$\pm$0.09
 &  8.20$\pm$0.10&  8.47$\pm$0.10 \\
5$\le$Events$\le$55$\,$MeV (\%)&  6.89$\pm$0.09&  6.98$\pm$0.09
 &  8.86$\pm$0.10&  9.11$\pm$0.11 \\

\hline
\multicolumn{1}{c}{$E_V$$\le$5$\,$MeV} 
&\multicolumn{1}{c}{$a_{z1,z2}$=0} 
&\multicolumn{1}{c}{{\tt TOMOGRAPHY} $a_{z1,z2}$} 
&\multicolumn{1}{c}{$a_{z1,z2}$=0} 
&\multicolumn{1}{c}{{\tt TOMOGRAPHY} $a_{z1,z2}$} \\

Peak Position (MeV)         & 68.88$\pm$0.03& 68.92$\pm$0.03
 & 70.07$\pm$0.03& 69.99$\pm$0.03 \\
FWHM$_{(220)}$ (MeV)        &  3.67$\pm$0.03&  4.09$\pm$0.03
 &  3.99$\pm$0.03&  4.76$\pm$0.03 \\
5$\le$Events$\le$54$\,$MeV (\%)&  1.91$\pm$0.05&  2.02$\pm$0.05
 &  3.92$\pm$0.07&  4.15$\pm$0.07 \\
5$\le$Events$\le$55$\,$MeV (\%)&  2.24$\pm$0.05&  2.36$\pm$0.05
 &  4.46$\pm$0.07&  4.68$\pm$0.07 \\
\hline
\end{tabular}
\end{table}\clearpage

\bigskip
\begin{table}[!tbp]
\caption{The predicted energy resolutions and tail contributions for
69.8$\,$MeV $e^+$\/ and 70.8$\,$MeV $\gamma$'s events in the PIBETA clusters
containing the crystal with maximum energy deposition and its nearest 
neighbors.}
\label{tab4}
\begin{tabular}{lcccc}
\hline
\multicolumn{1}{c}{Parameter} 
&\multicolumn{1}{c}{69.8$\,$MeV $e^+$}
& 
&\multicolumn{1}{c}{70.8$\,$MeV $\gamma$} 
& \\ 

\multicolumn{1}{c}{No $E_V$\/ cut} 
&\multicolumn{1}{c}{$a_{z1,z2}$=0} 
&\multicolumn{1}{c}{{\tt TOMOGRAPHY} $a_{z1,z2}$} 
&\multicolumn{1}{c}{$a_{z1,z2}$=0} 
&\multicolumn{1}{c}{{\tt TOMOGRAPHY} $a_{z1,z2}$} \\
\hline\hline
Peak Position (MeV)         & 67.74$\pm$0.03& 67.85$\pm$0.03
 & 68.80$\pm$0.03& 68.65$\pm$0.03 \\
FWHM$_{\rm (NN)}$ (MeV)     &  5.22$\pm$0.03&  5.90$\pm$0.03
 &  6.01$\pm$0.03&  6.80$\pm$0.03 \\
5$\le$Events$\le$54$\,$MeV (\%)&  3.83$\pm$0.07&  4.01$\pm$0.07
 &  5.47$\pm$0.08&  5.78$\pm$0.08 \\
5$\le$Events$\le$55$\,$MeV (\%)&  4.54$\pm$0.07&  4.73$\pm$0.08
 &  6.33$\pm$0.09&  6.67$\pm$0.09 \\

\hline
\multicolumn{1}{c}{$E_V$$\le$5$\,$MeV} 
&\multicolumn{1}{c}{$a_{z1,z2}$=0} 
&\multicolumn{1}{c}{{\tt TOMOGRAPHY} $a_{z1,z2}$} 
&\multicolumn{1}{c}{$a_{z1,z2}$=0} 
&\multicolumn{1}{c}{{\tt TOMOGRAPHY} $a_{z1,z2}$} \\

Peak Position (MeV)         & 67.75$\pm$0.03& 67.83$\pm$0.03
 & 68.80$\pm$0.04& 68.64$\pm$0.03 \\
FWHM$_{\rm (NN)}$ (MeV)     &  5.52$\pm$0.03&  5.83$\pm$0.03
 &  5.92$\pm$0.03&  6.76$\pm$0.03 \\
5$\le$Events$\le$54$\,$MeV (\%)&  3.53$\pm$0.07&  3.73$\pm$0.07
 &  5.28$\pm$0.08&  5.59$\pm$0.08 \\
5$\le$Events$\le$55$\,$MeV (\%)&  4.19$\pm$0.07&  4.39$\pm$0.07
 &  6.10$\pm$0.09&  6.42$\pm$0.09 \\
\hline
\end{tabular}
\end{table}
\clearpage

\vspace*{\stretch{1}}
\centerline{\psfig{figure=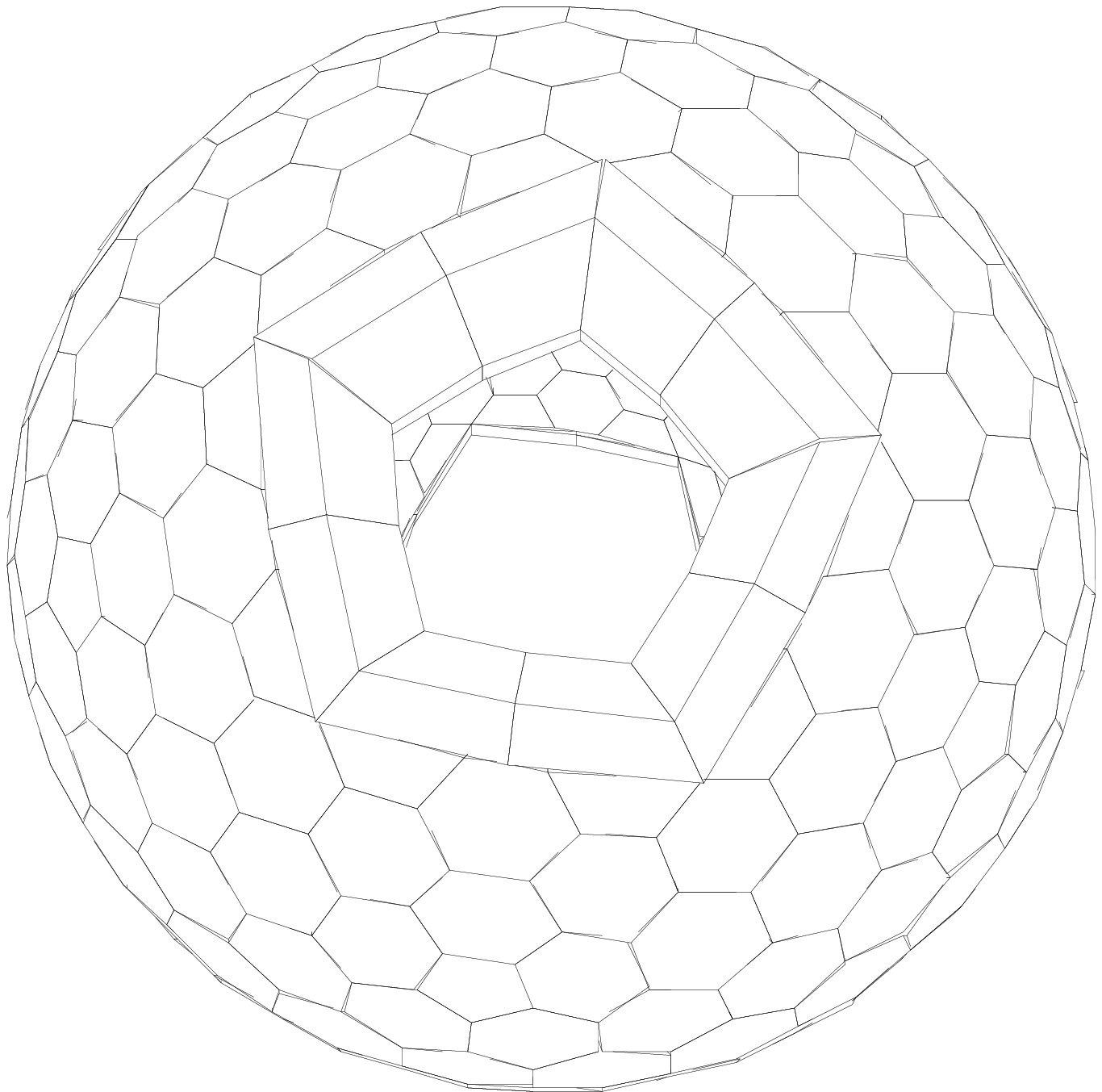,width=10cm}}
\bigskip\bigskip\bigskip
\centerline{FIGURE 1}
\vspace*{\stretch{2}}
\clearpage

\vspace*{\stretch{1}}
\centerline{\psfig{figure=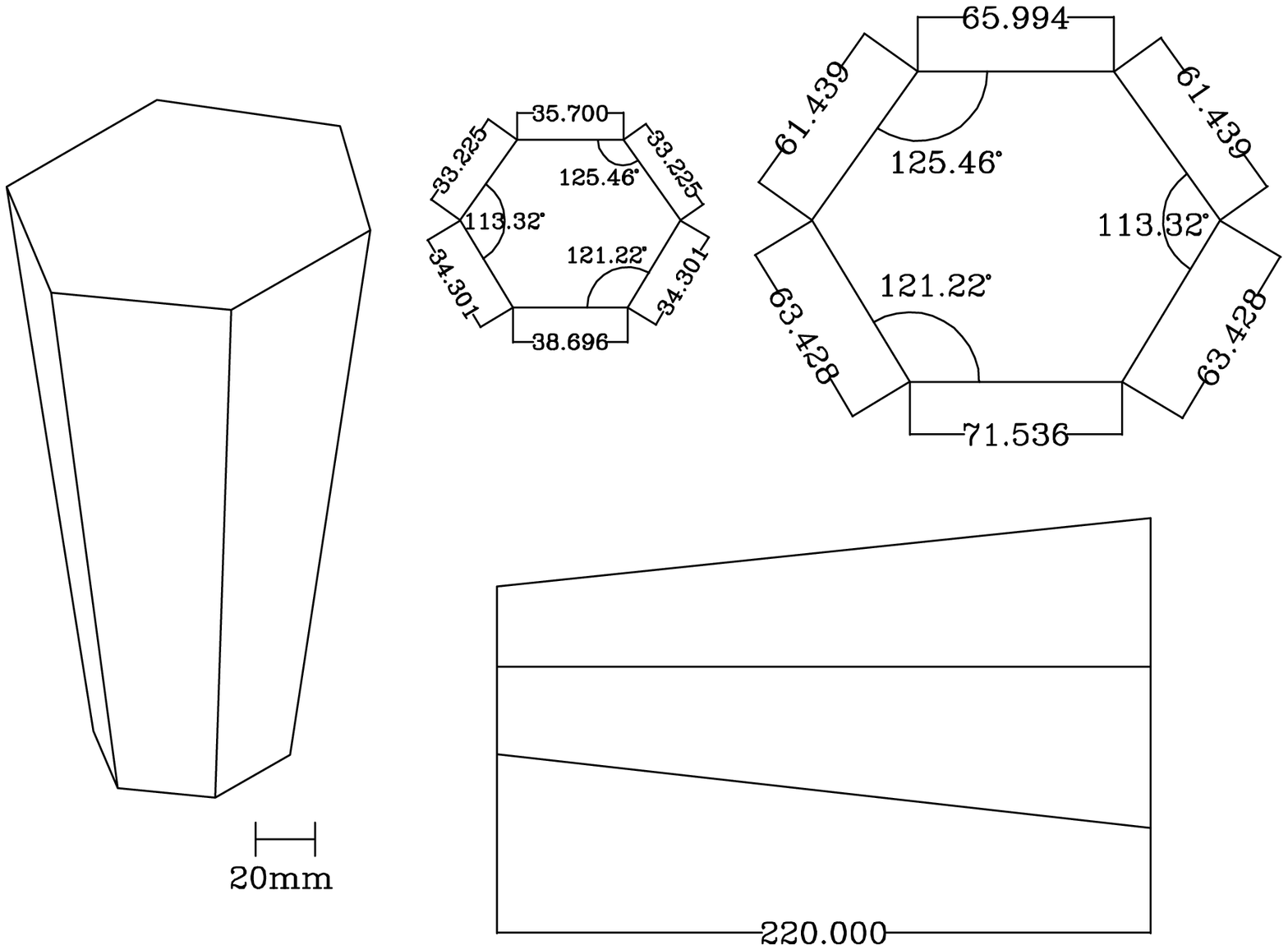,width=12cm}}
\bigskip\bigskip
\centerline{\psfig{figure=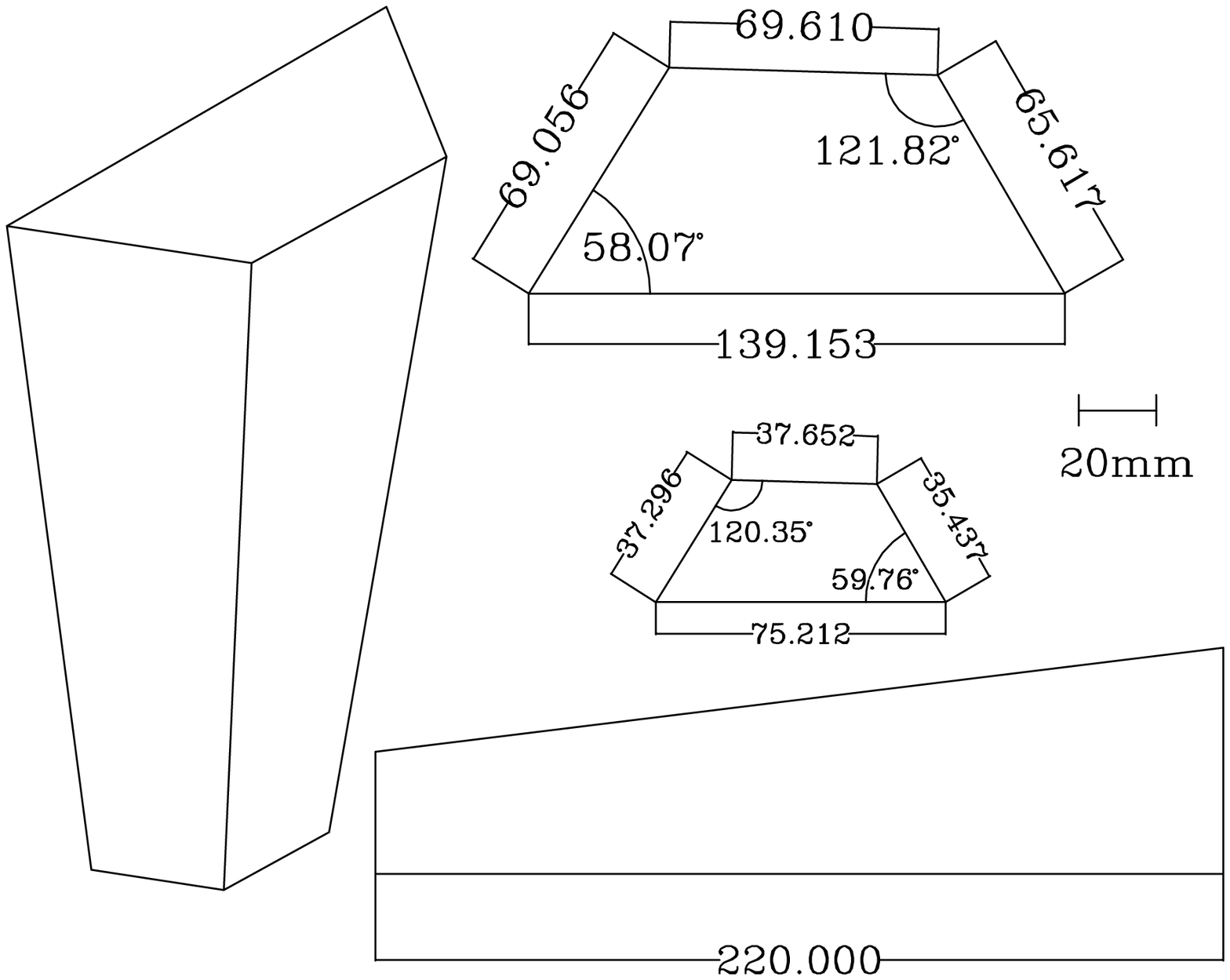,width=12cm}}
\vglue 2cm
\centerline{FIGURE 2}
\vspace*{\stretch{2}}
\clearpage

\vspace*{\stretch{1}}
\centerline{\psfig{figure=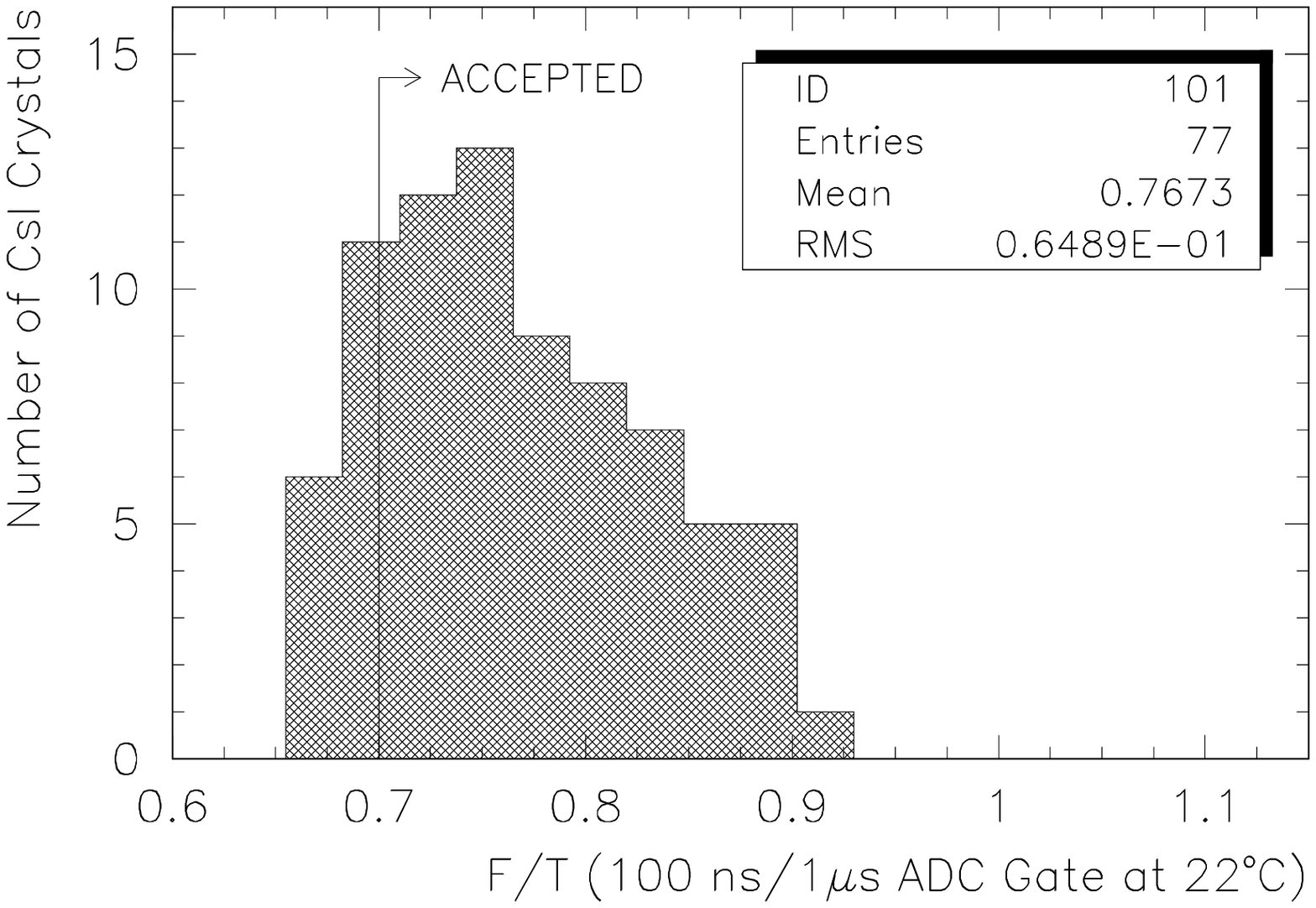,height=22cm}}
\vglue -9cm
\centerline{FIGURE 3}
\vspace*{\stretch{2}}
\clearpage

\vspace*{\stretch{1}}
\centerline{\psfig{figure=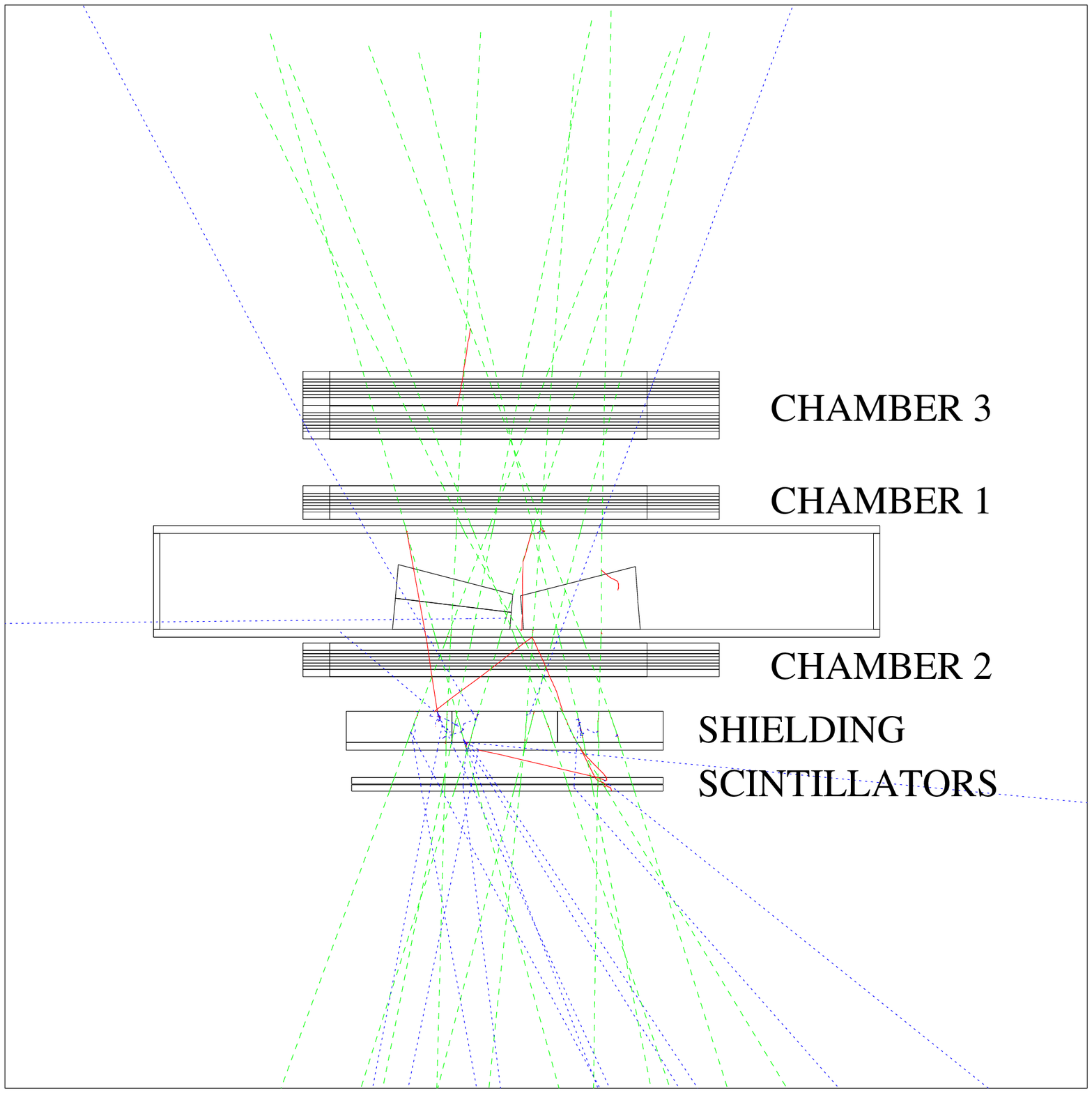,width=12.7cm}}
\vglue -1cm
\centerline{FIGURE 4}
\vspace*{\stretch{2}}
\clearpage

\vspace*{\stretch{1}}
\centerline{\psfig{figure=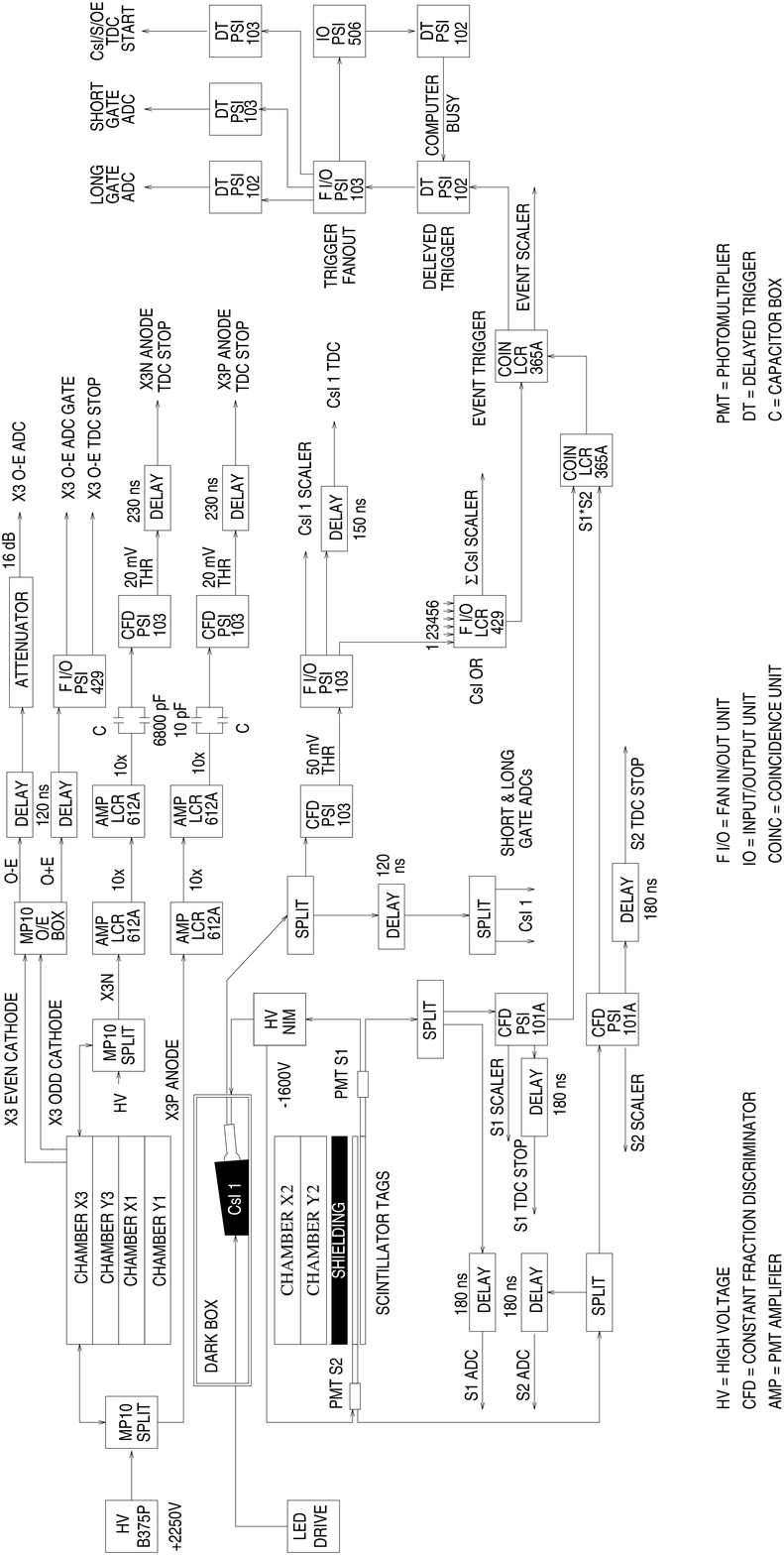,width=10cm}}
\bigskip\bigskip
\centerline{FIGURE 5}
\vspace*{\stretch{2}}
\clearpage

\vspace*{\stretch{1}}
\centerline{\psfig{figure=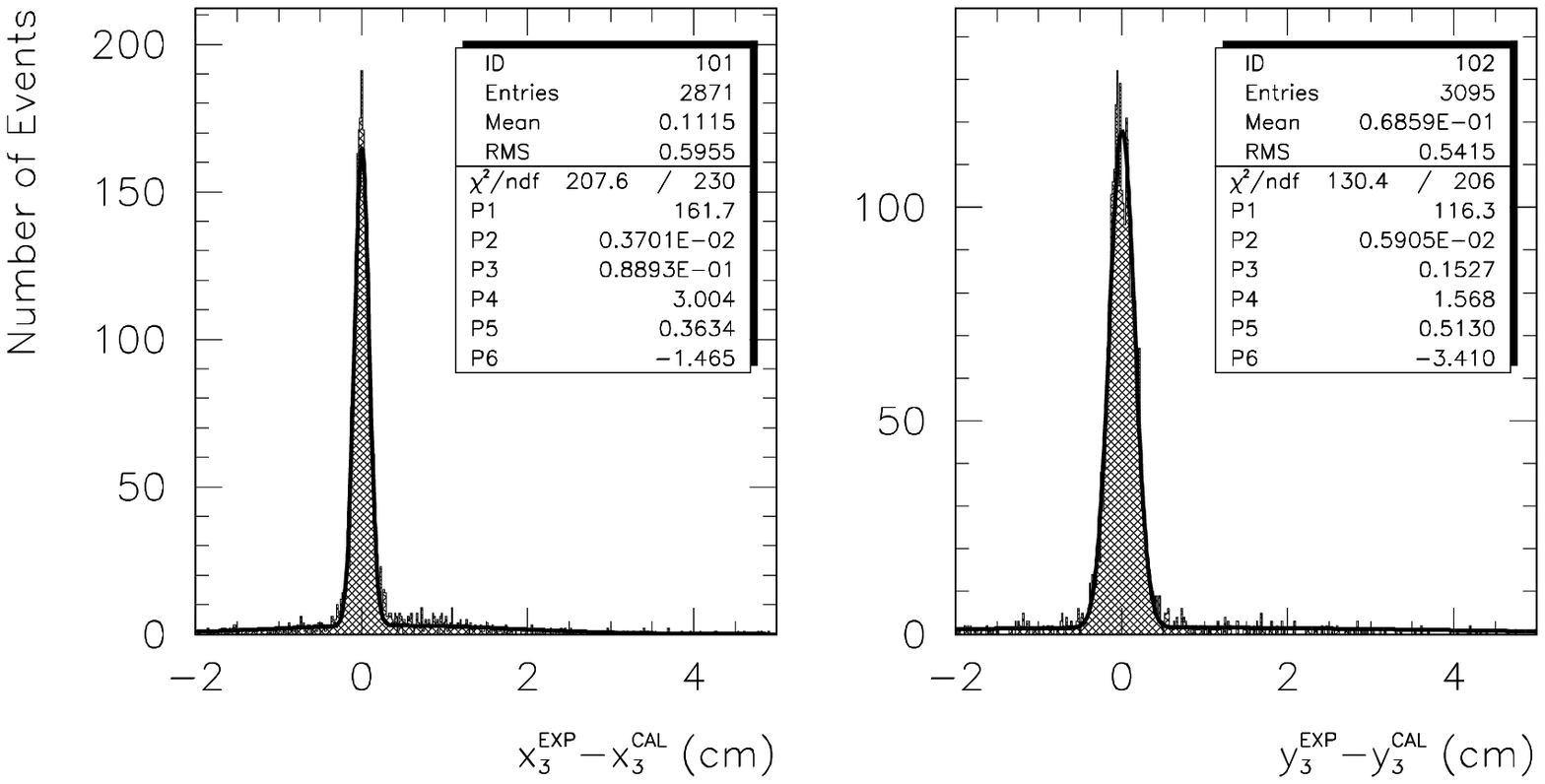,height=20cm}}
\vglue -8cm
\centerline{FIGURE 6}
\vspace*{\stretch{2}}
\clearpage

\vspace*{\stretch{1}}
\centerline{\psfig{figure=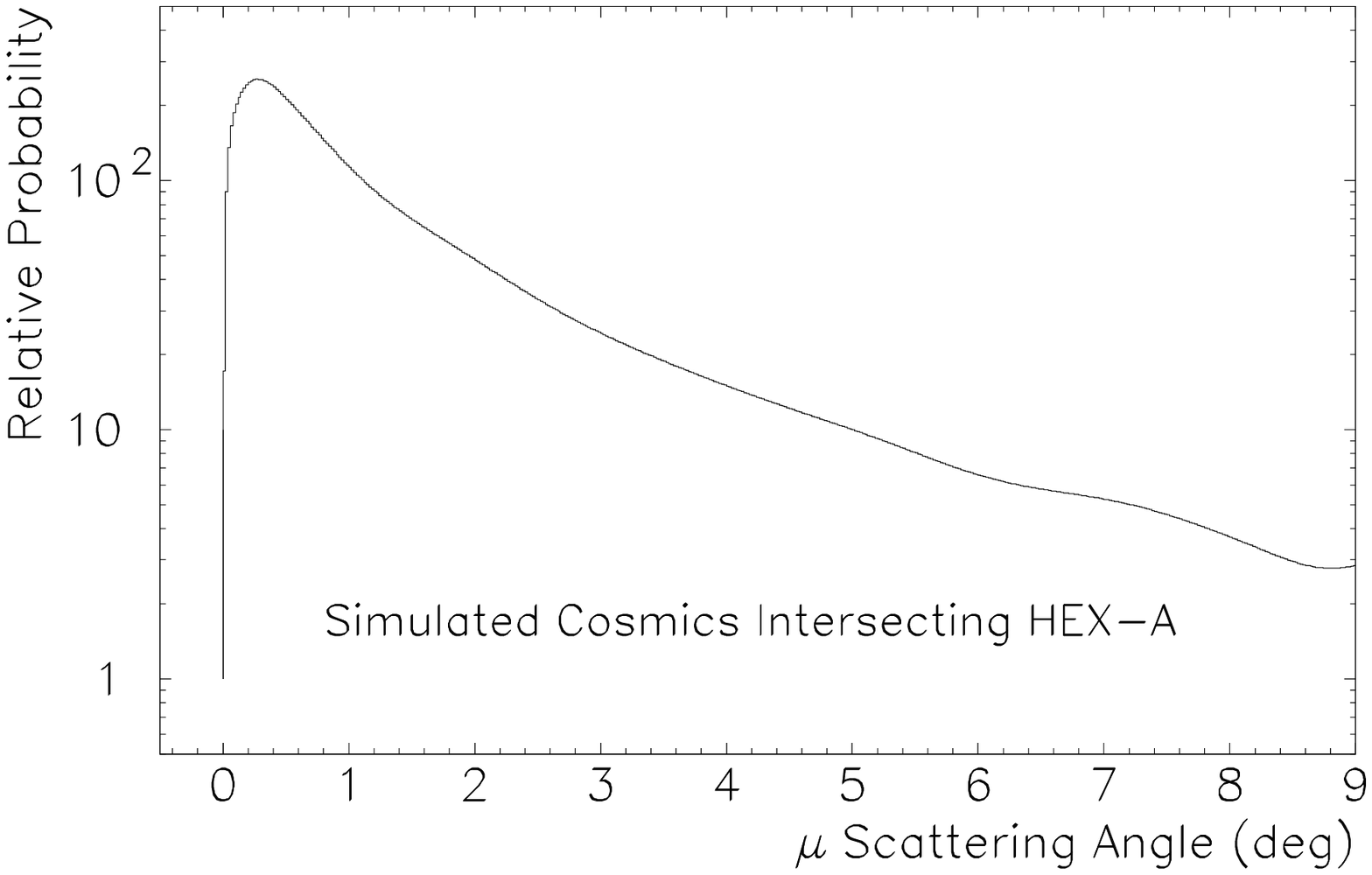,height=22cm}}
\vglue -10cm
\centerline{FIGURE 7}
\vspace*{\stretch{2}}
\clearpage

\vspace*{\stretch{1}}
\centerline{\psfig{figure=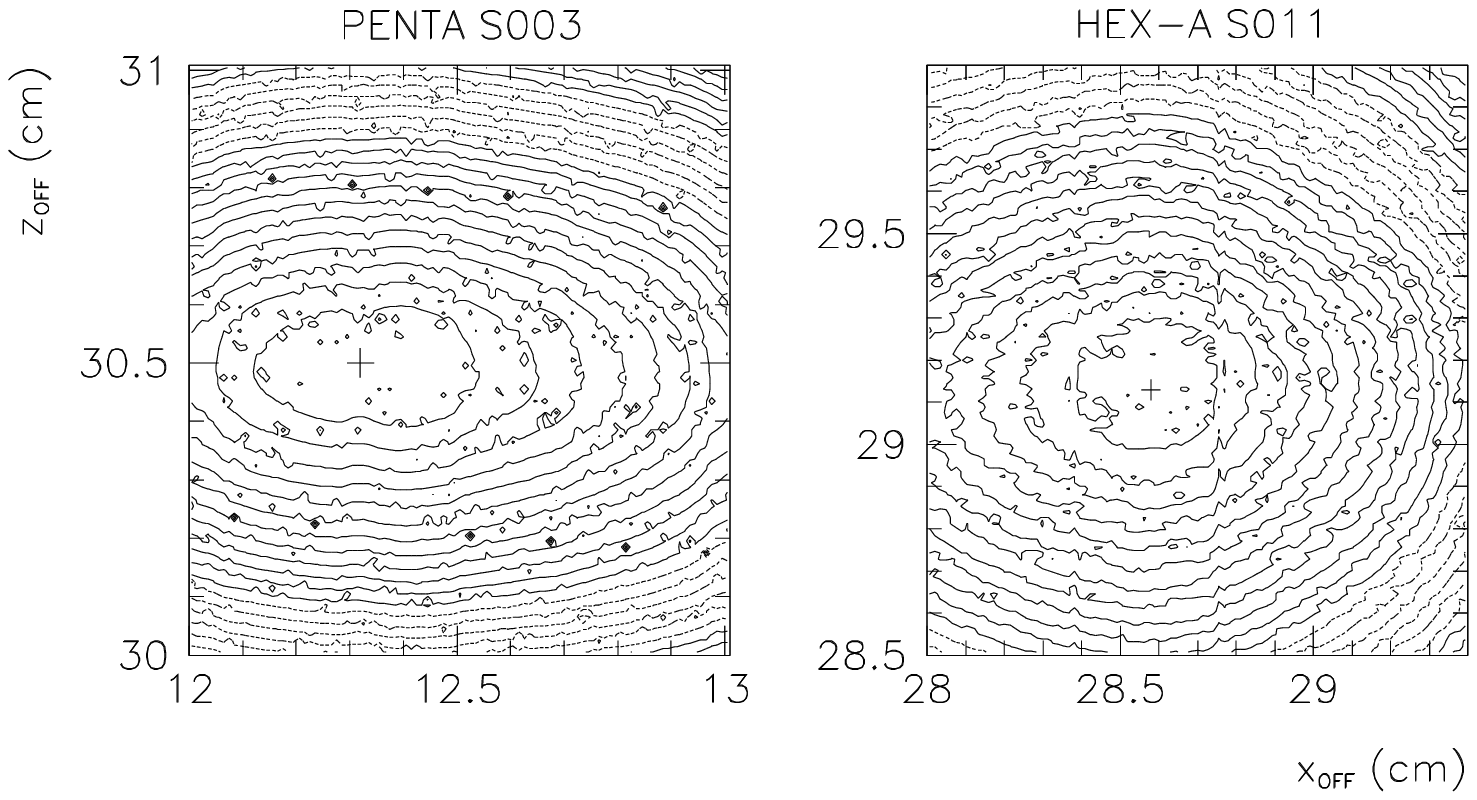,width=20cm}}
\vglue -9cm
\centerline{FIGURE 8}
\vspace*{\stretch{2}}
\clearpage

\vspace*{\stretch{1}}
\centerline{\psfig{figure=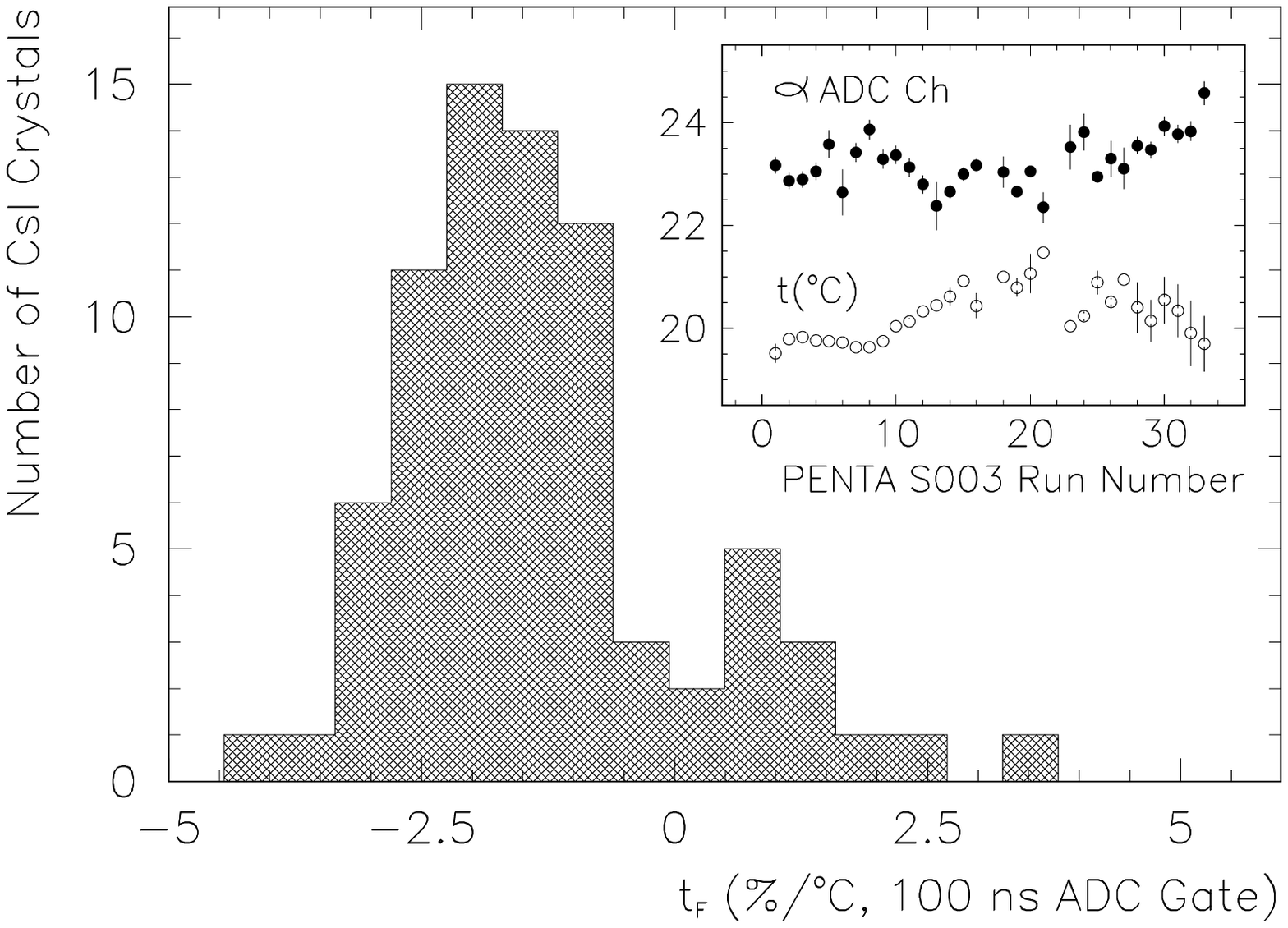,height=22.0cm}}
\vglue -9cm
\centerline{FIGURE 9}
\vspace*{\stretch{2}}
\clearpage

\vspace*{\stretch{1}}
\centerline{\psfig{figure=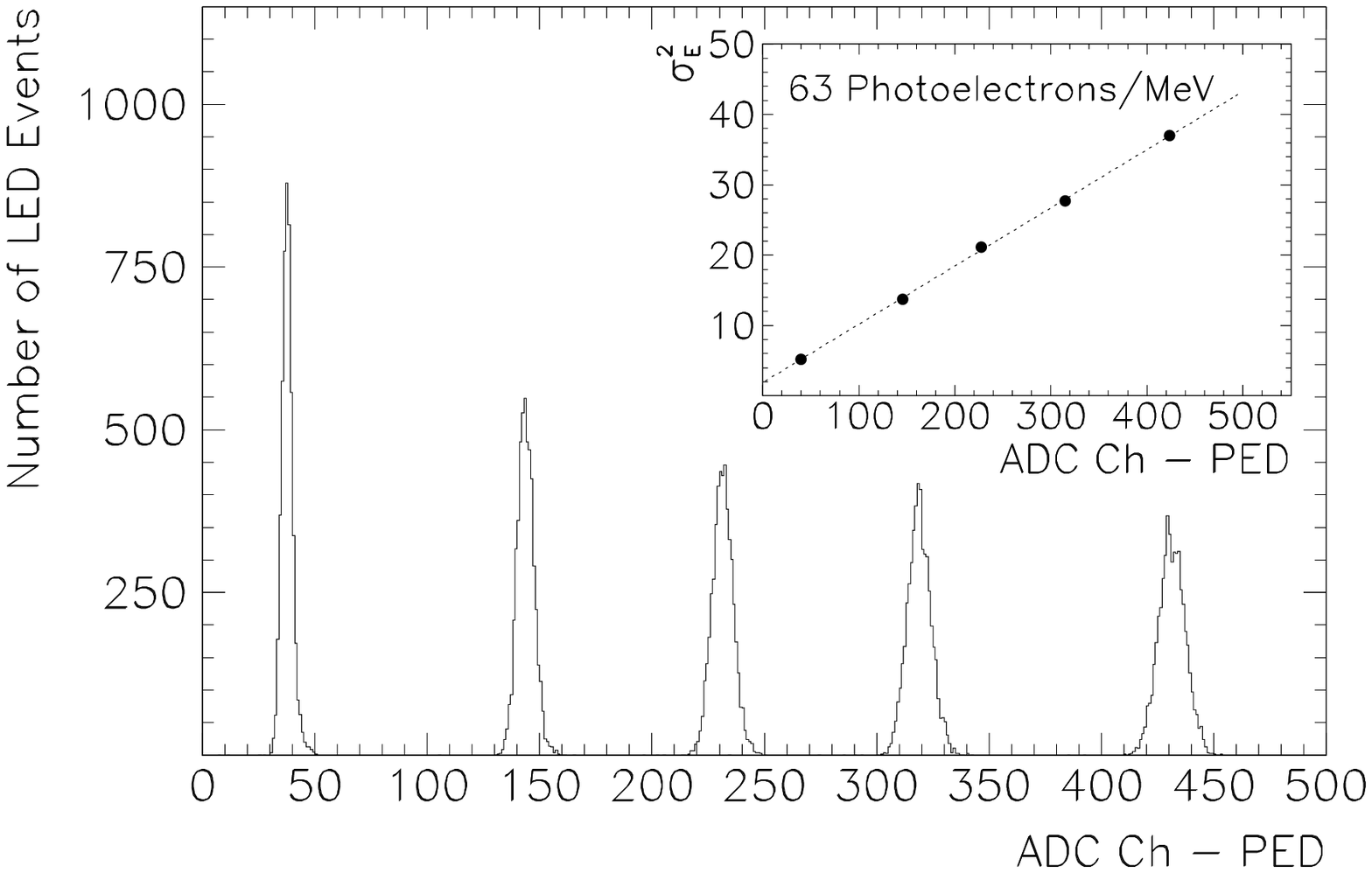,height=22cm}}
\vglue -9cm
\centerline{FIGURE 10}
\vspace*{\stretch{2}}
\clearpage

\vspace*{\stretch{1}}
\centerline{\psfig{figure=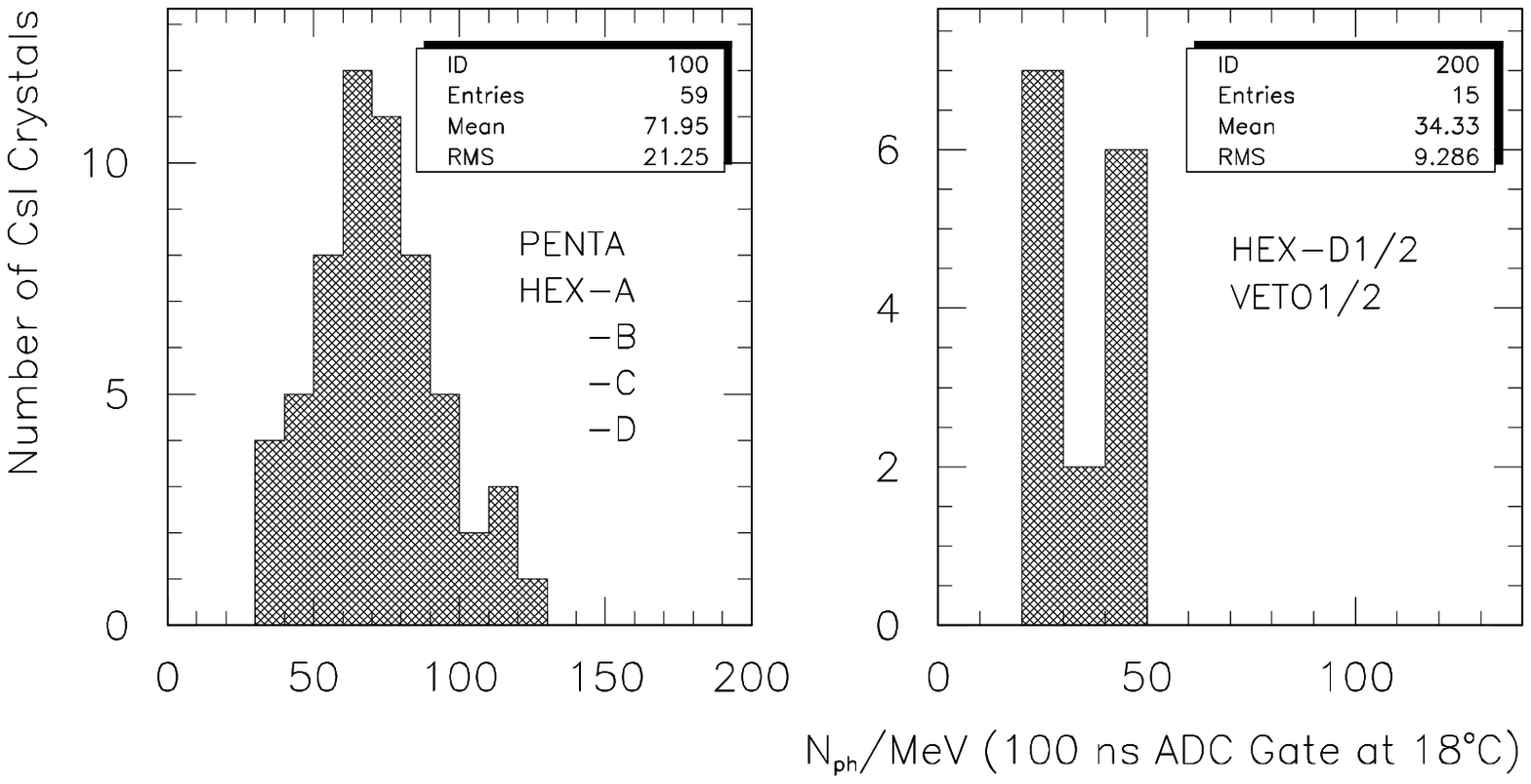,height=22cm}}
\vglue -9cm
\centerline{FIGURE 11}
\vspace*{\stretch{2}}
\clearpage

\vspace*{\stretch{1}}
\centerline{\psfig{figure=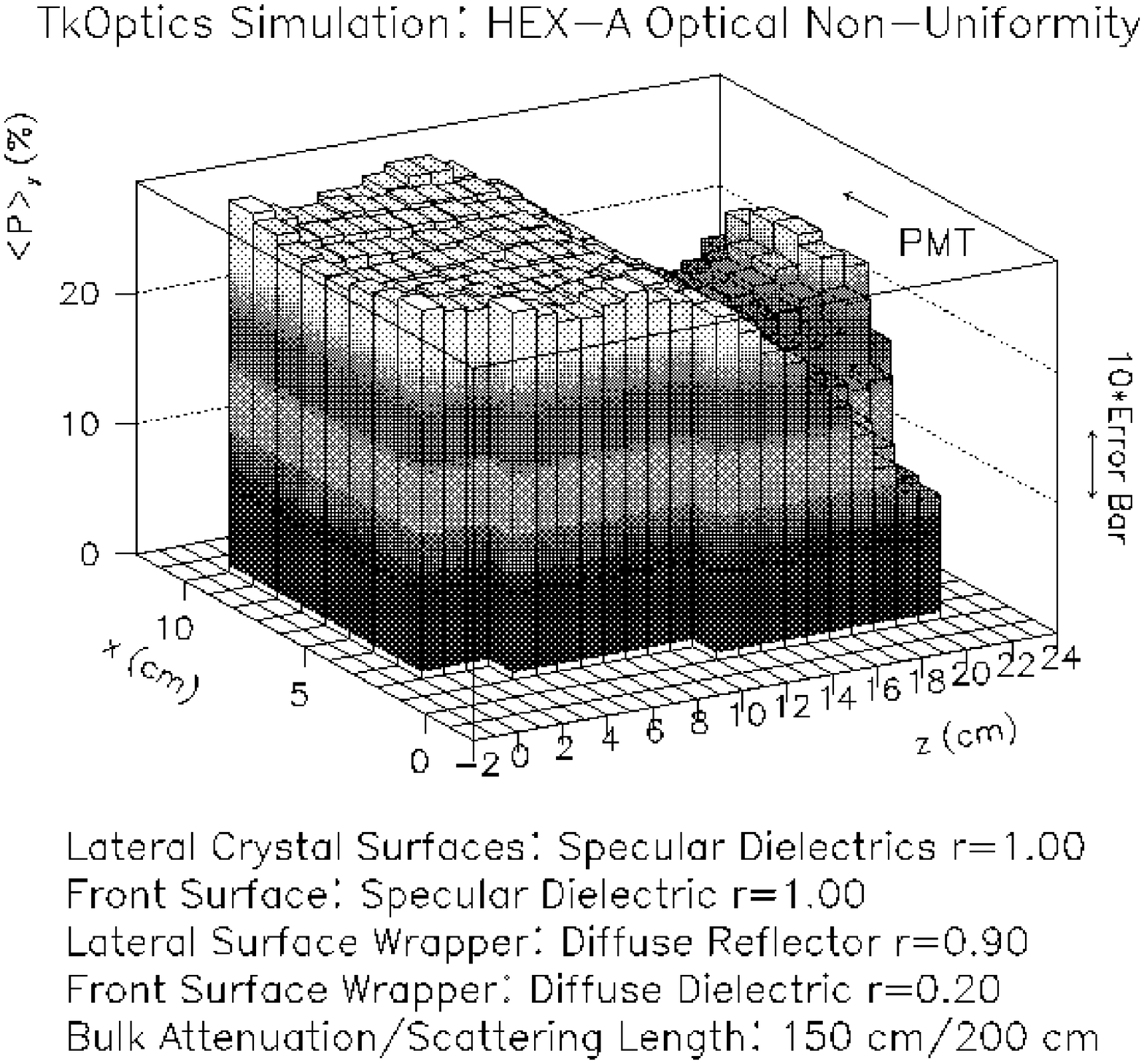,width=15cm}}
\vglue -4.5cm
\centerline{FIGURE 12}
\vspace*{\stretch{2}}
\clearpage

\vspace*{\stretch{1}}
\centerline{\psfig{figure=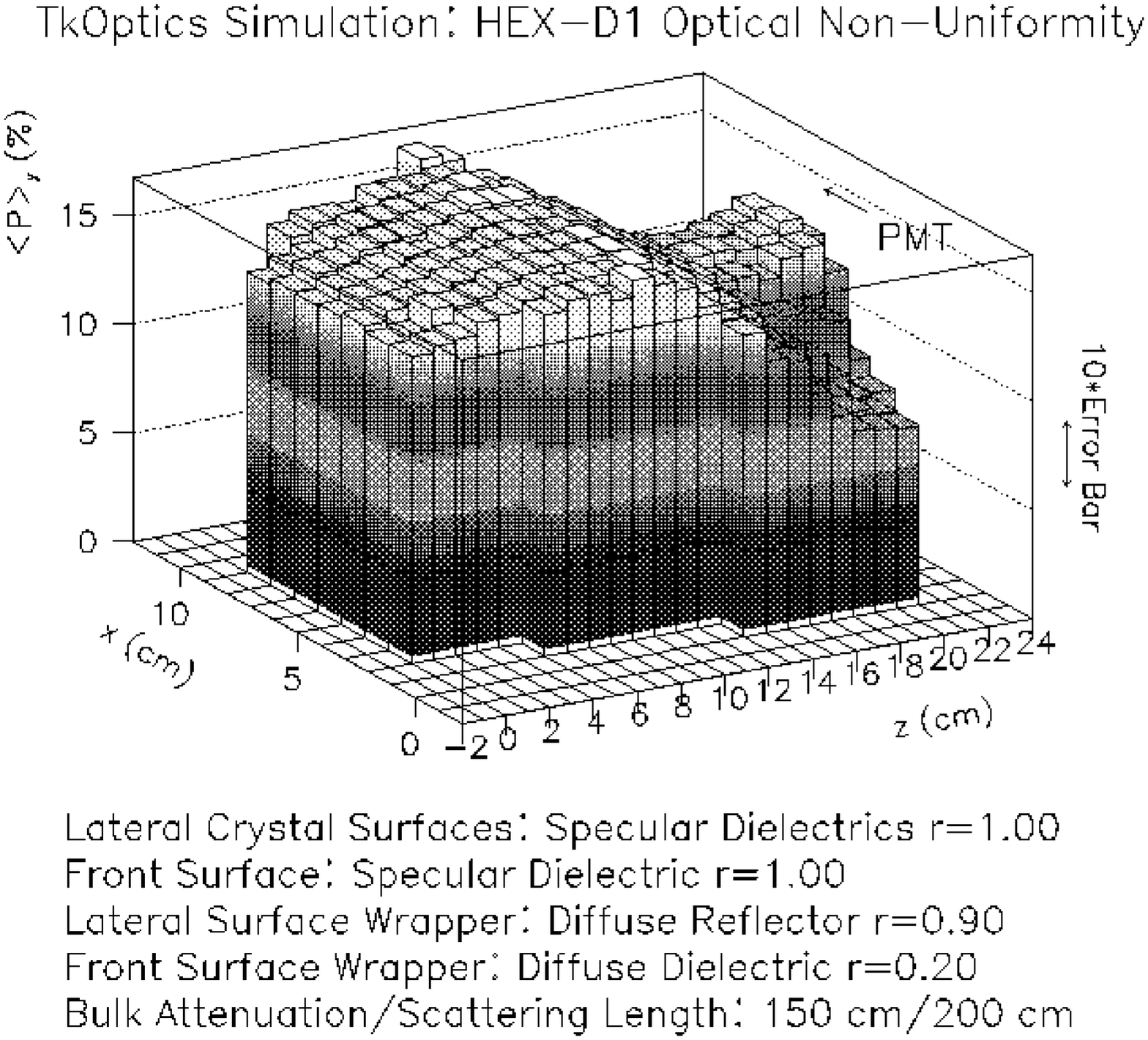,width=15cm}}
\vglue -4.5cm
\centerline{FIGURE 13}
\vspace*{\stretch{2}}
\clearpage

\vspace*{\stretch{1}}
\vglue -1cm
\centerline{\psfig{figure=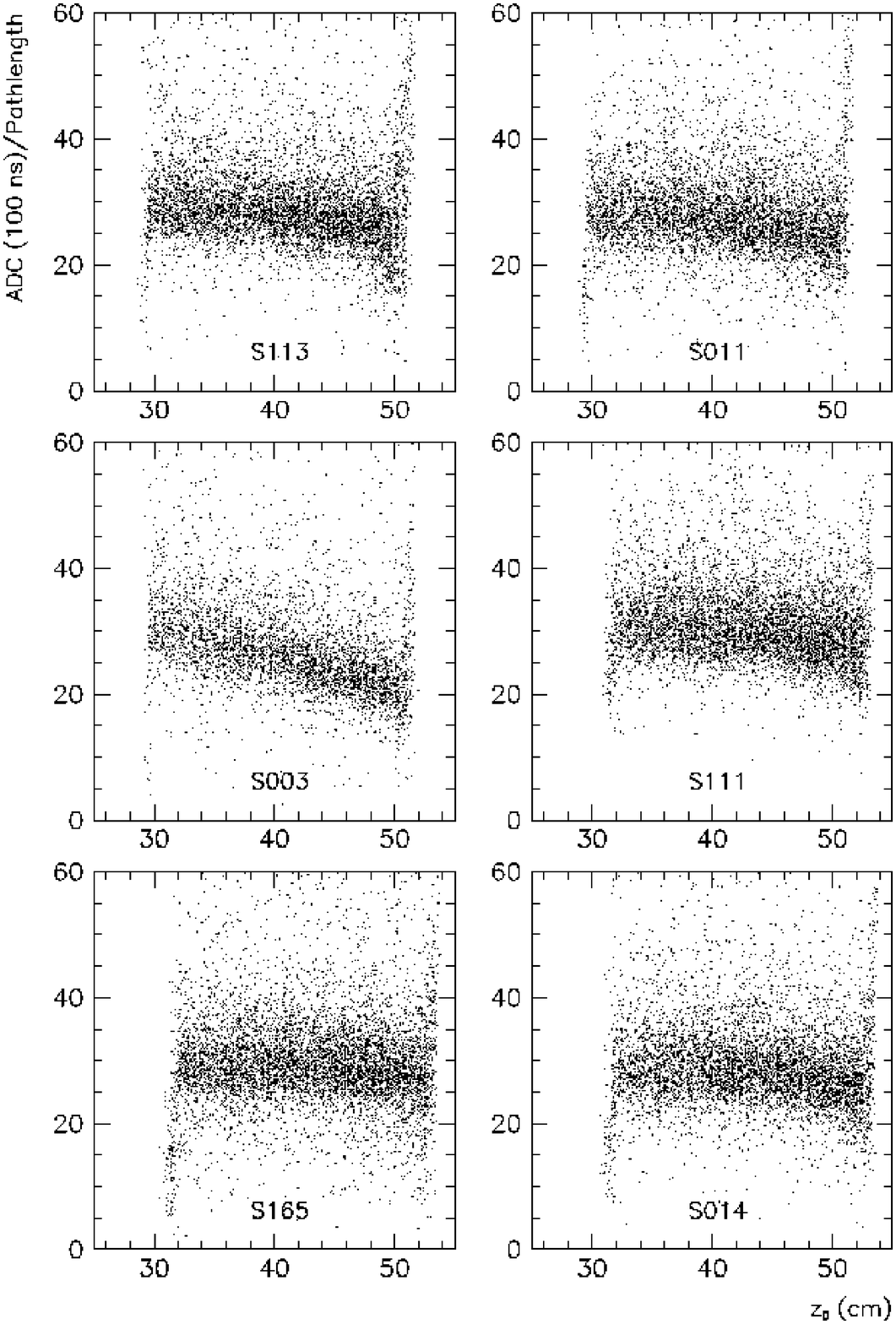,height=22cm}}
\vglue 0.5cm
\centerline{FIGURE 14}
\vspace*{\stretch{2}}
\clearpage

\vspace*{\stretch{1}}
\vglue -1cm
\centerline{\psfig{figure=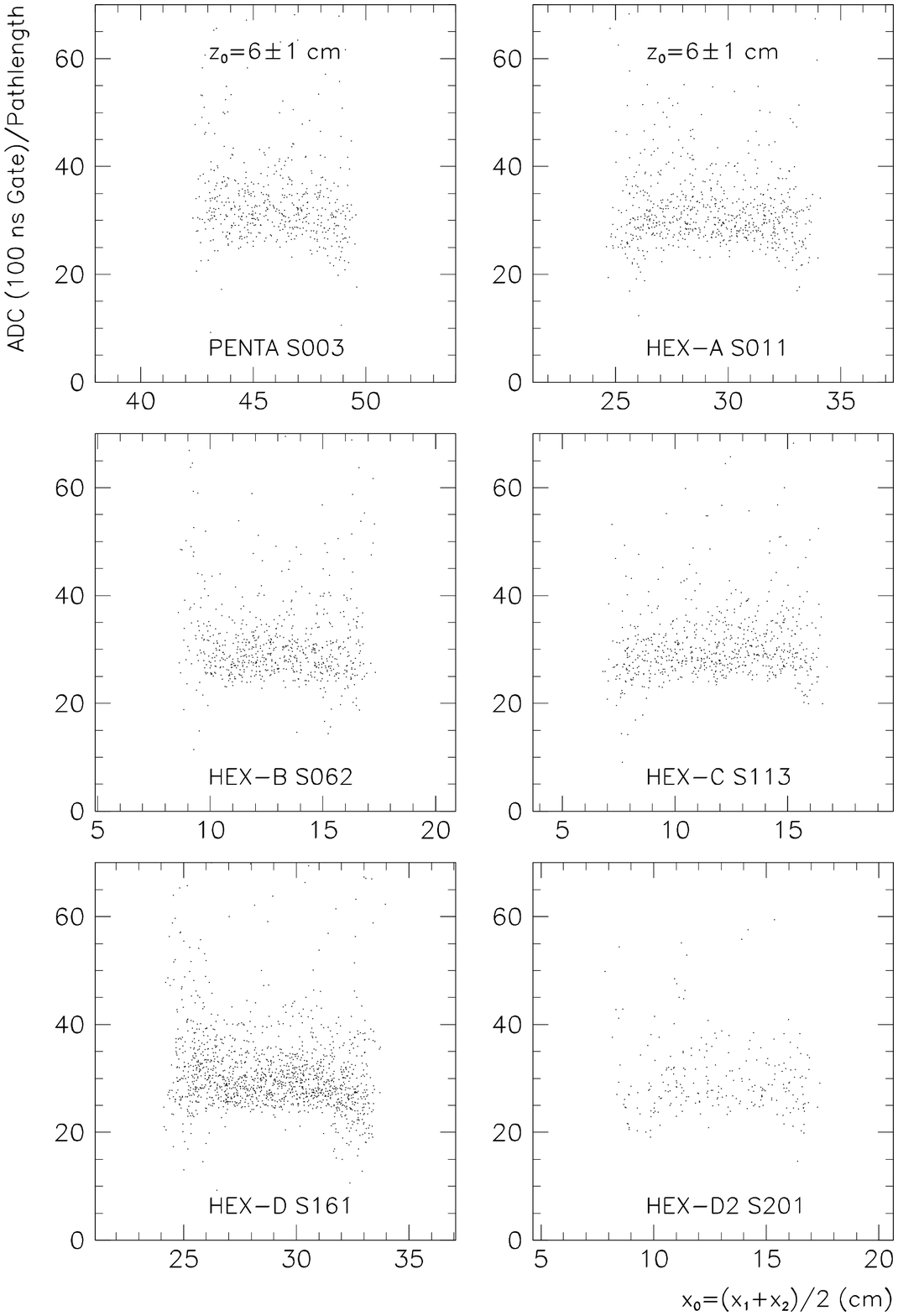,height=22cm}}
\vglue 0.5cm
\centerline{FIGURE 15}
\vspace*{\stretch{2}}
\clearpage

\vspace*{\stretch{1}}
\vglue -1cm
\centerline{\psfig{figure=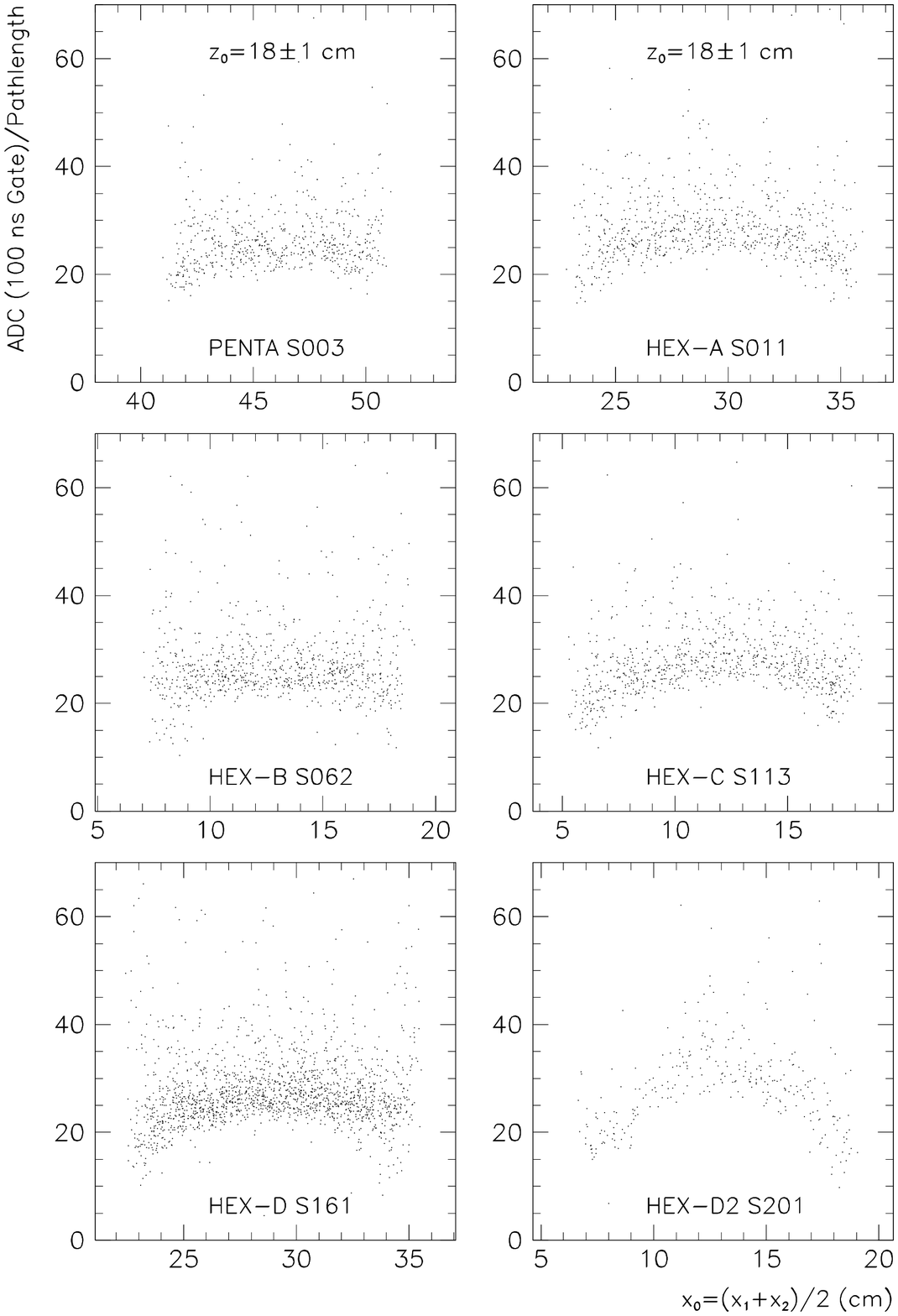,height=22cm}}
\vglue 0.5cm
\centerline{FIGURE 16}
\vspace*{\stretch{2}}
\clearpage

\vspace*{\stretch{1}}
\centerline{\psfig{figure=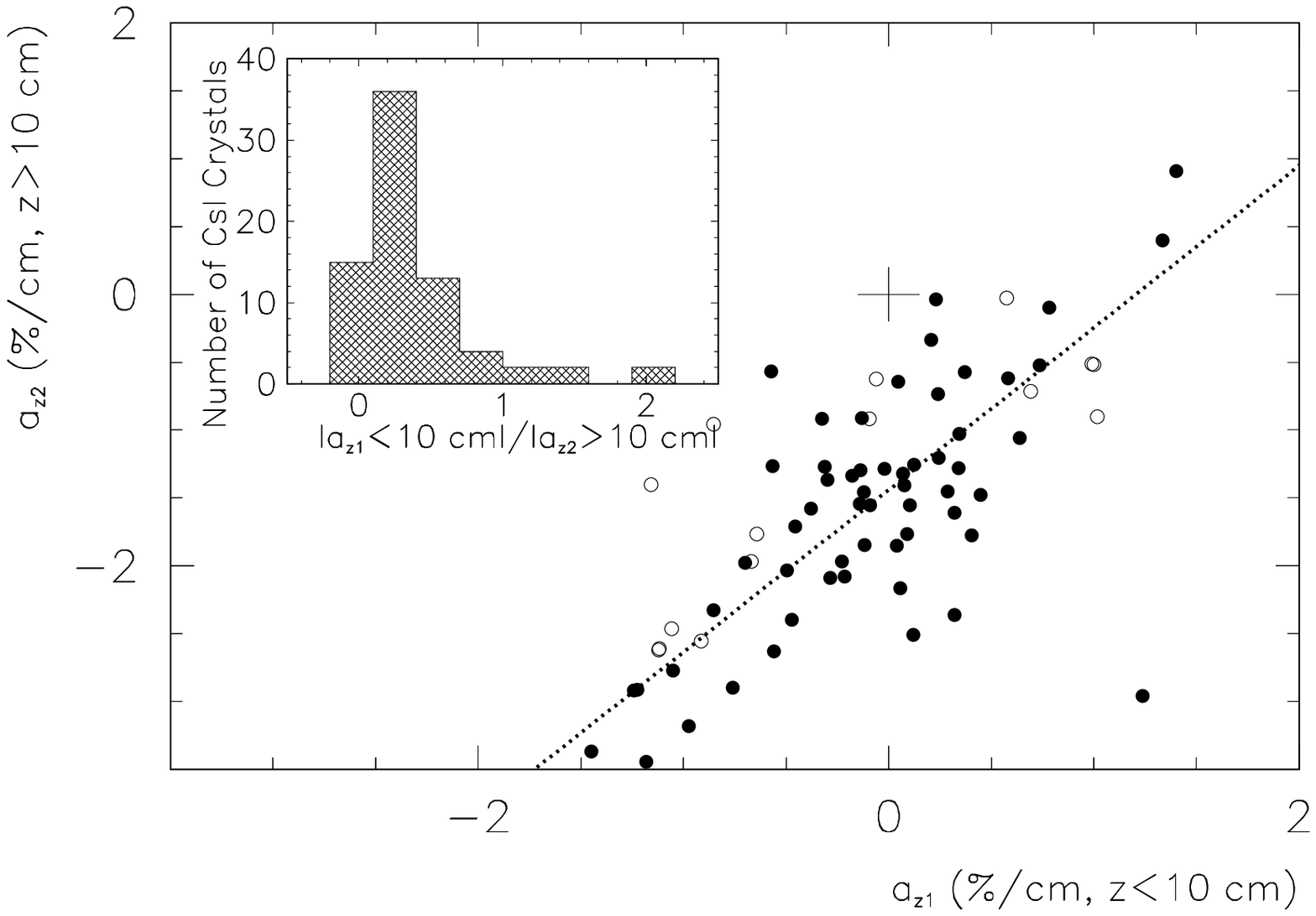,height=22cm}}
\vglue -9cm
\centerline{FIGURE 17}
\vspace*{\stretch{2}}
\clearpage

\vspace*{\stretch{1}}
\centerline{\psfig{figure=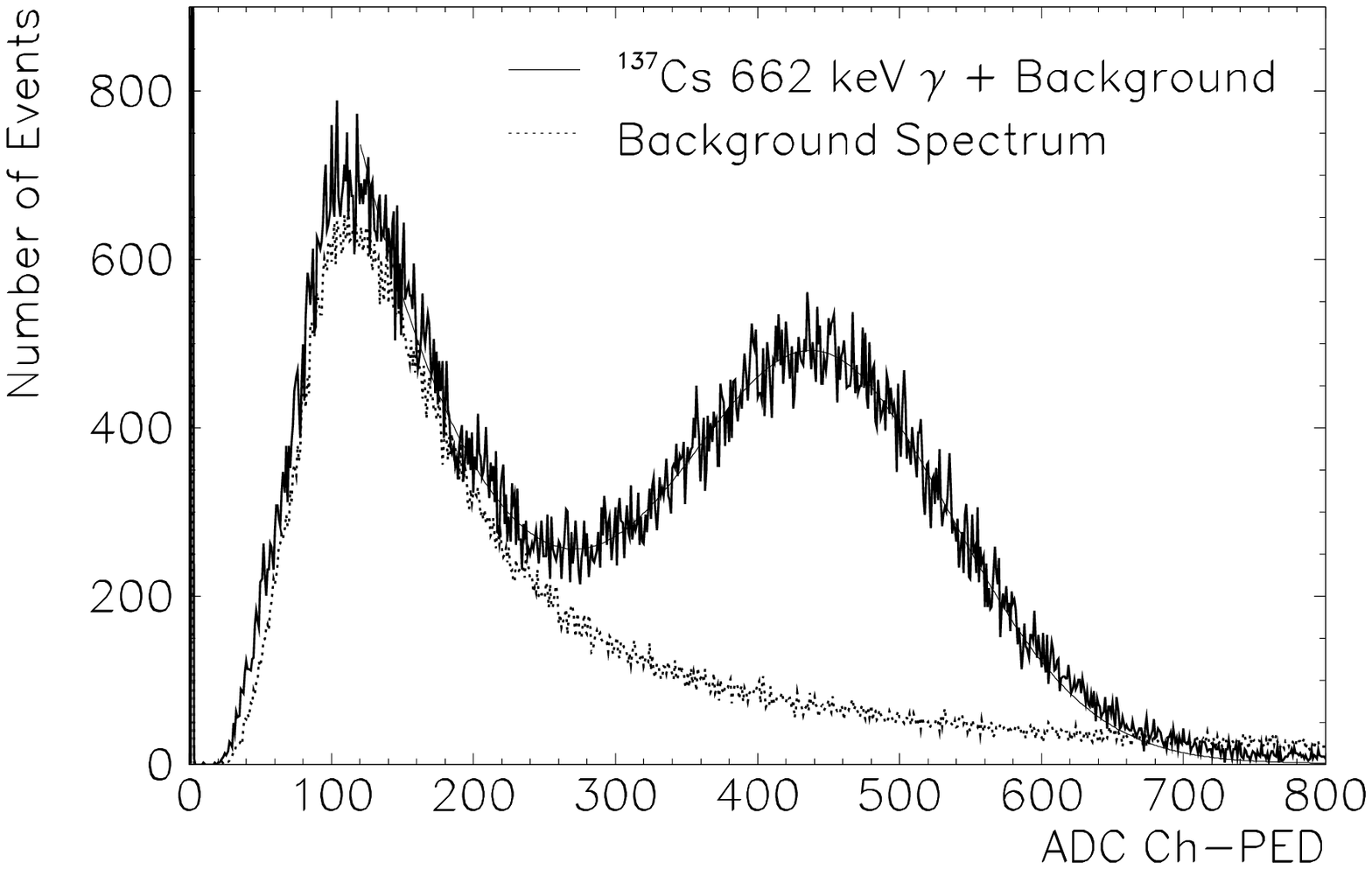,height=22cm}}
\vglue -9cm
\centerline{FIGURE 18}
\vspace*{\stretch{2}}
\clearpage

\vspace*{\stretch{1}}
\vglue -1cm
\centerline{\psfig{figure=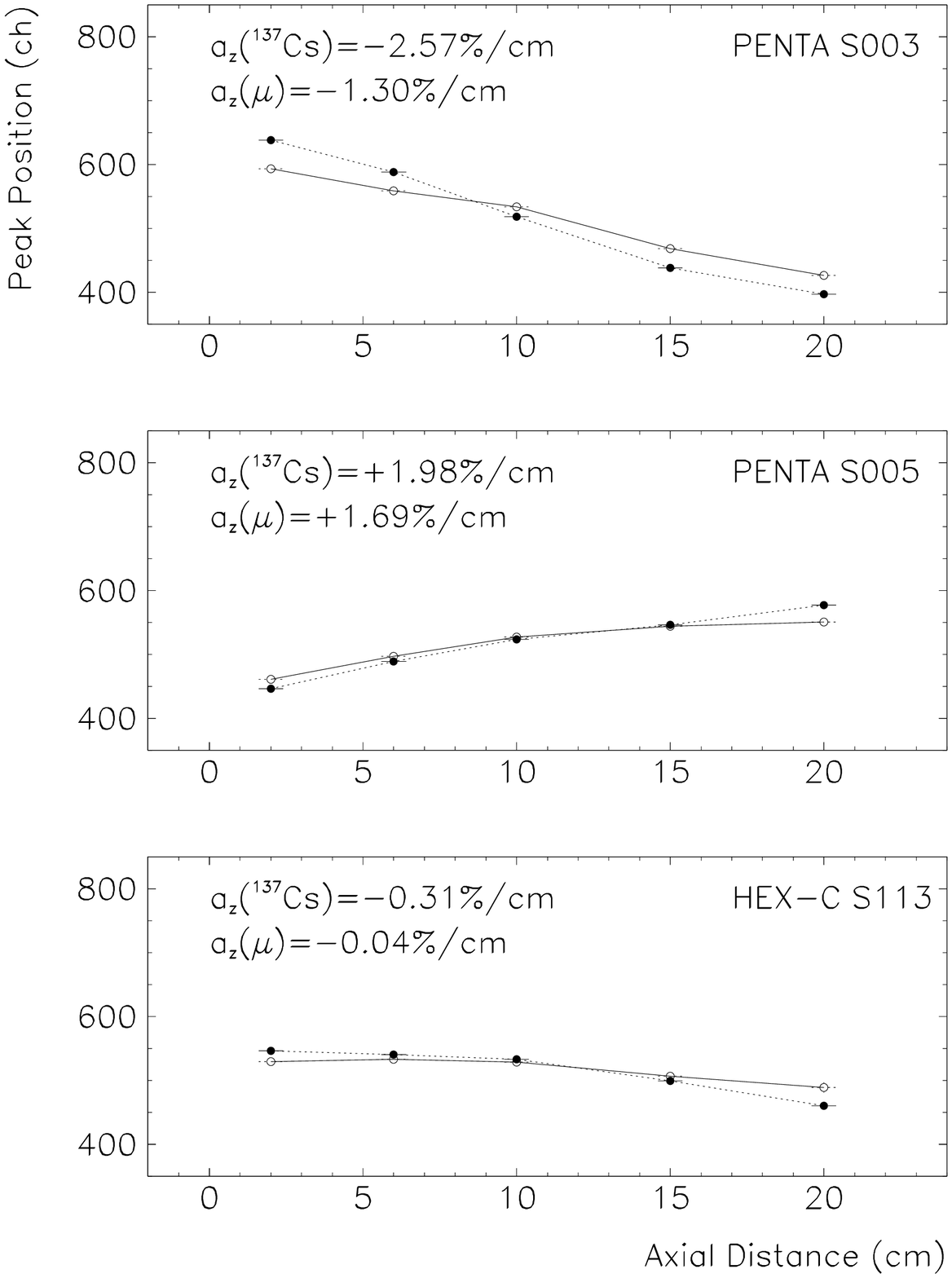,height=22cm}}
\vglue 0.5cm
\centerline{FIGURE 19}
\vspace*{\stretch{2}}
\clearpage

\vspace*{\stretch{1}}
\vglue -1cm
\centerline{\psfig{figure=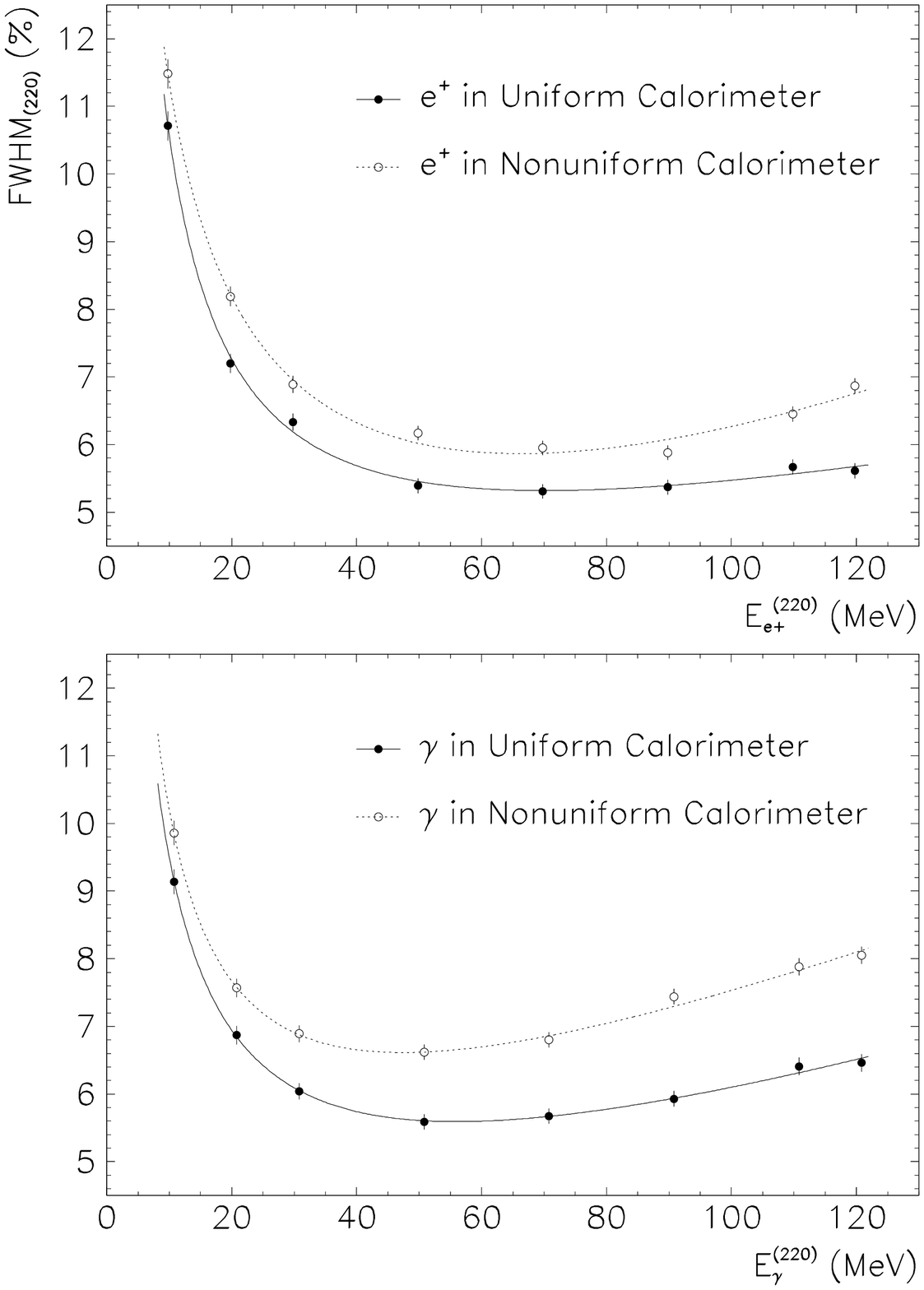,height=22cm}}
\vglue 0.5cm
\centerline{FIGURE 20}
\vspace*{\stretch{2}}
\clearpage

\vspace*{\stretch{1}}
\vglue -1cm
\centerline{\psfig{figure=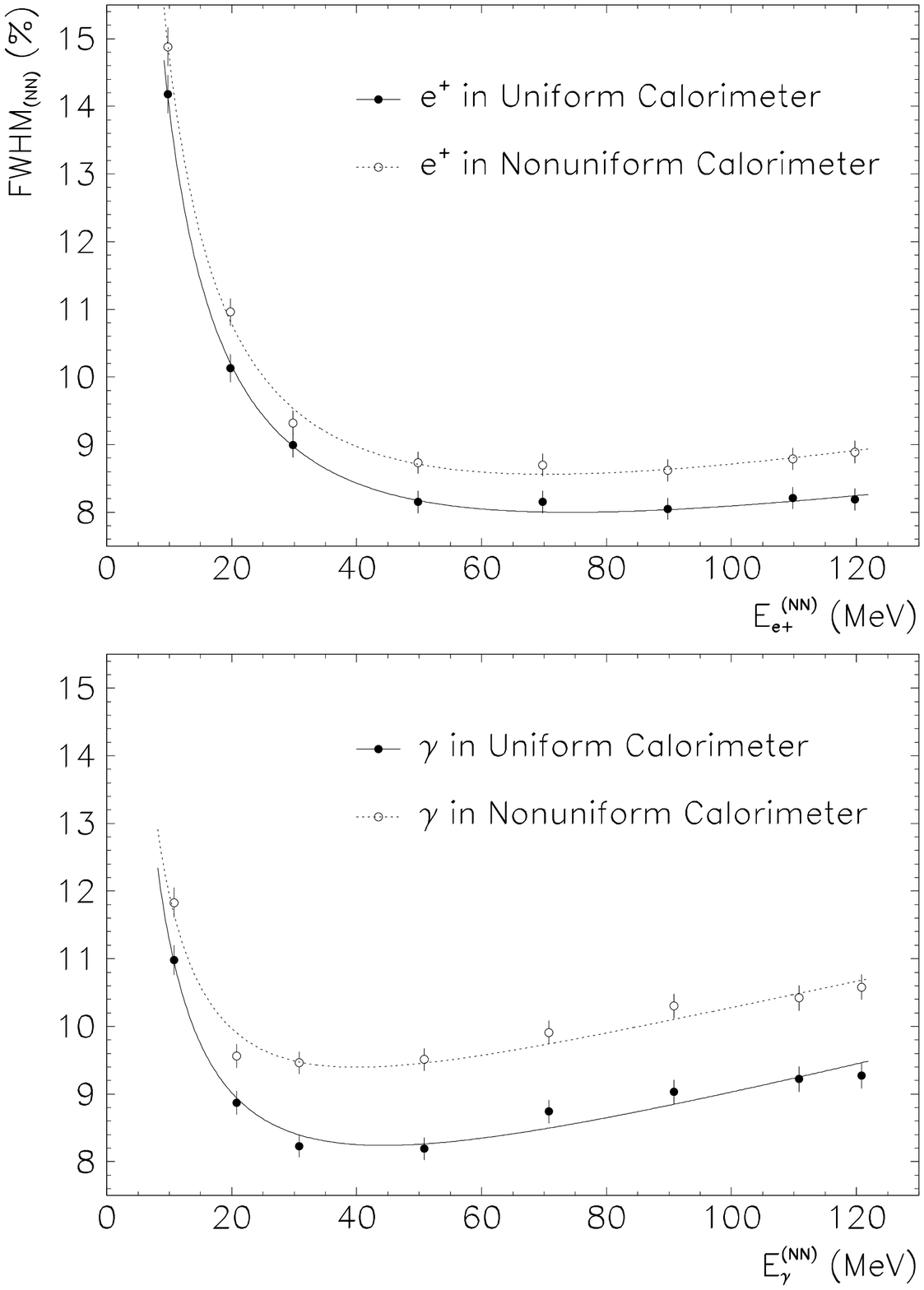,height=22cm}}
\vglue 0.5cm
\centerline{FIGURE 21}
\vspace*{\stretch{2}}
\clearpage

\vspace*{\stretch{1}}
\vglue -1cm
\centerline{\psfig{figure=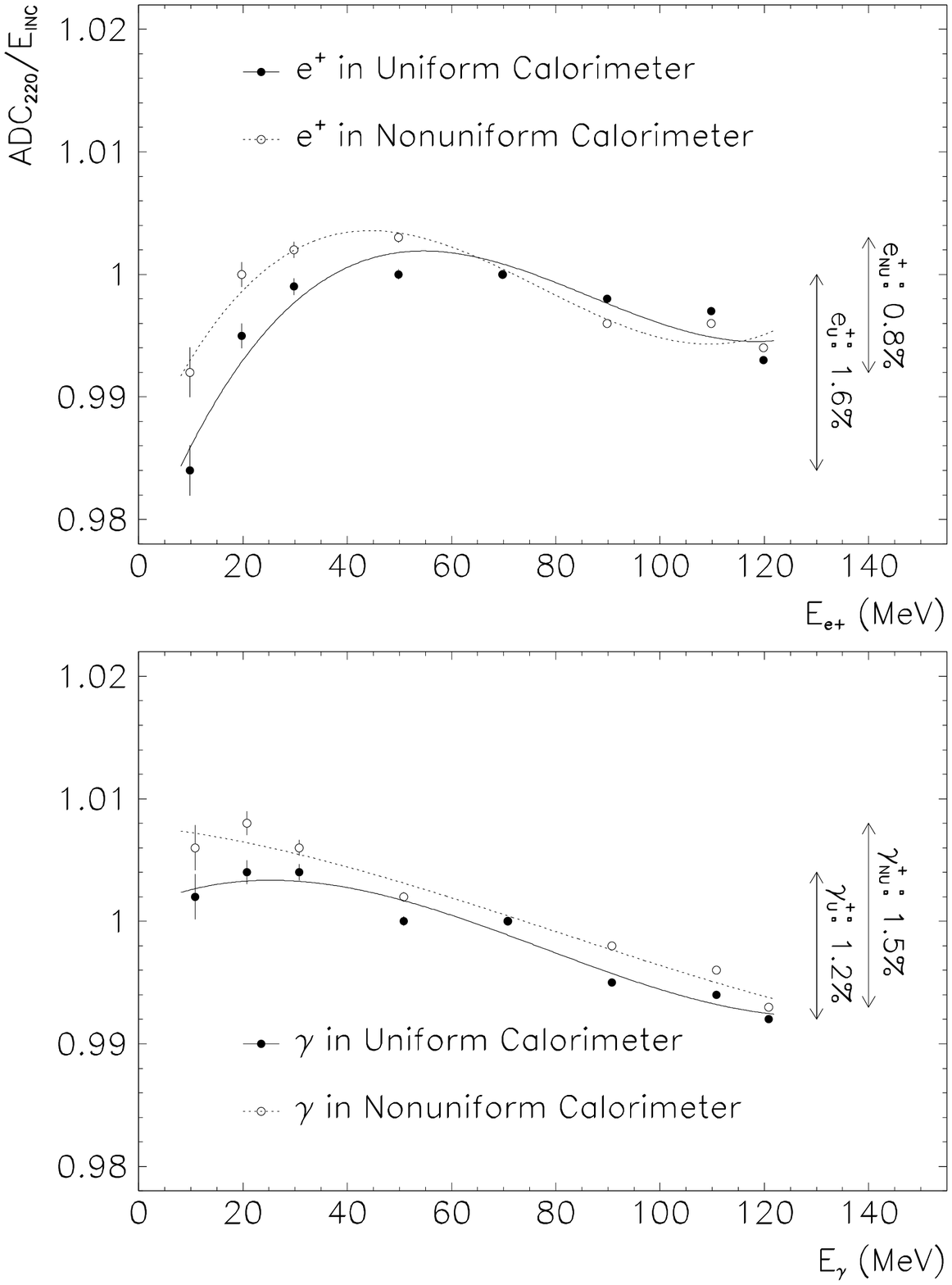,height=22cm}}
\vglue 0.5cm
\centerline{FIGURE 22}
\vspace*{\stretch{2}}
\clearpage

\vspace*{\stretch{1}}
\vglue -1cm
\centerline{\psfig{figure=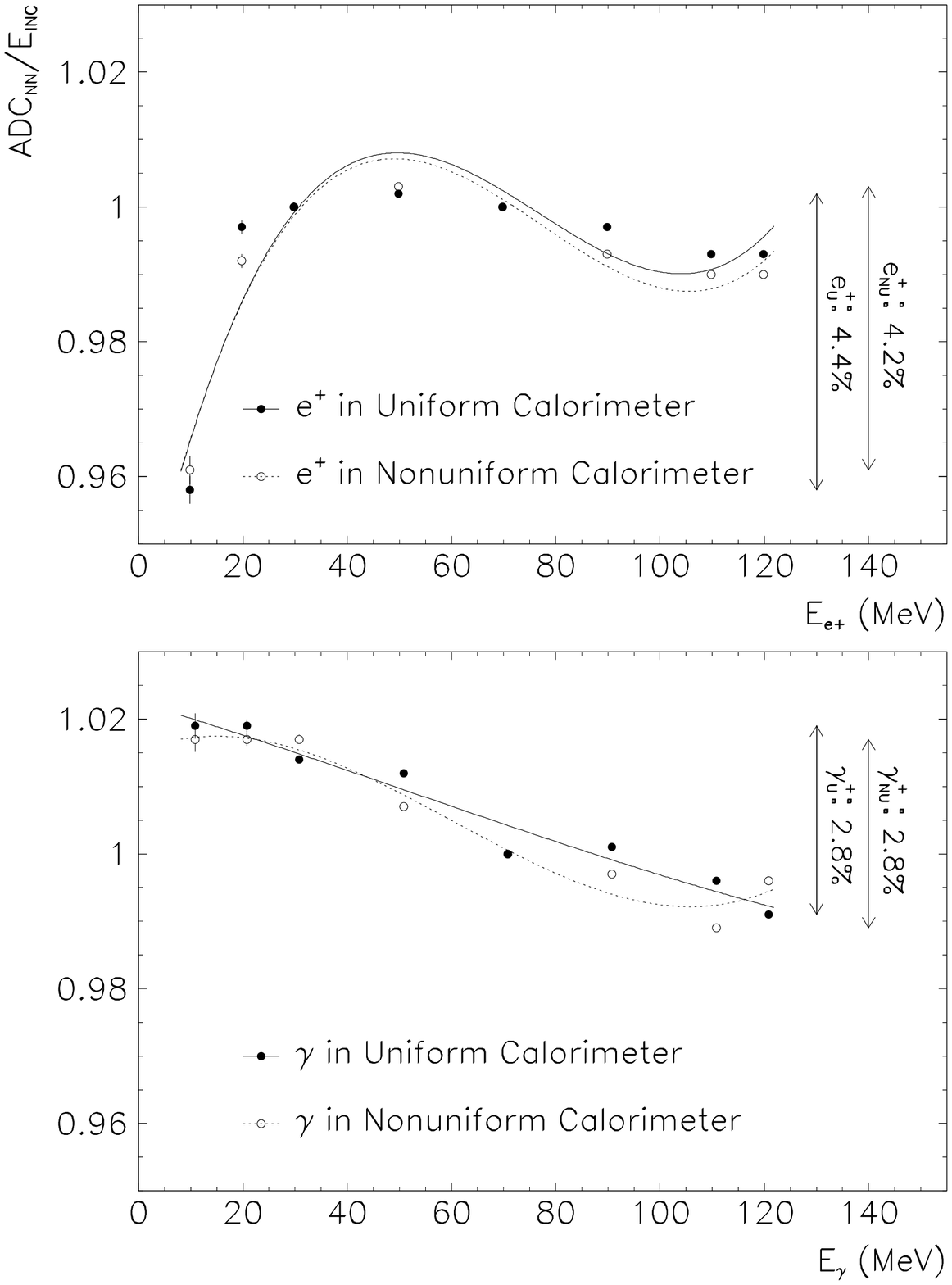,height=22cm}}
\vglue 0.5cm
\centerline{FIGURE 23}
\vspace*{\stretch{2}}
\clearpage

\vspace*{\stretch{1}}
\centerline{\psfig{figure=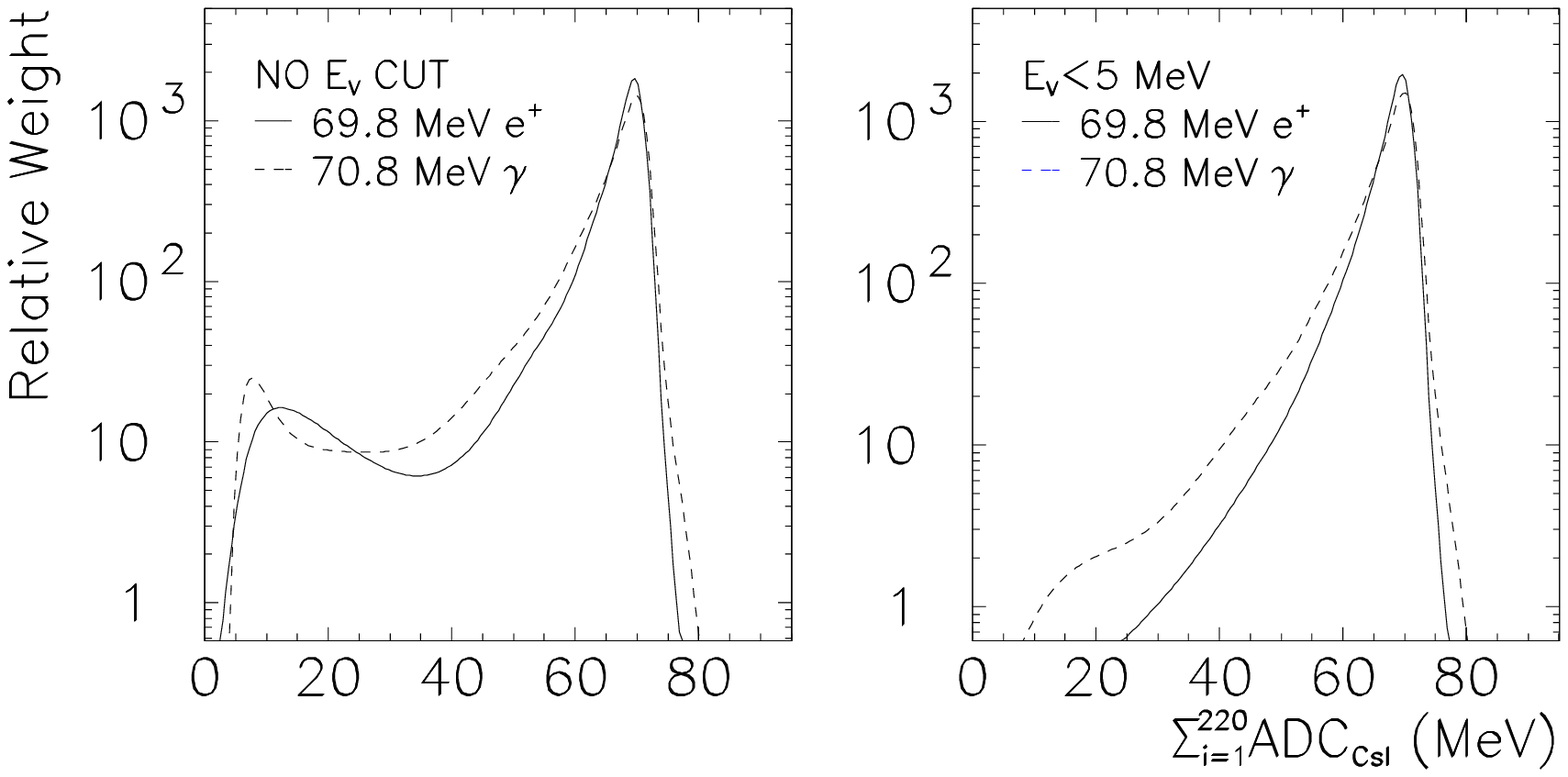,height=20cm}}
\vglue -8.5cm
\centerline{FIGURE 24}
\vspace*{\stretch{2}}
\clearpage

\vspace*{\stretch{1}}
\centerline{\psfig{figure=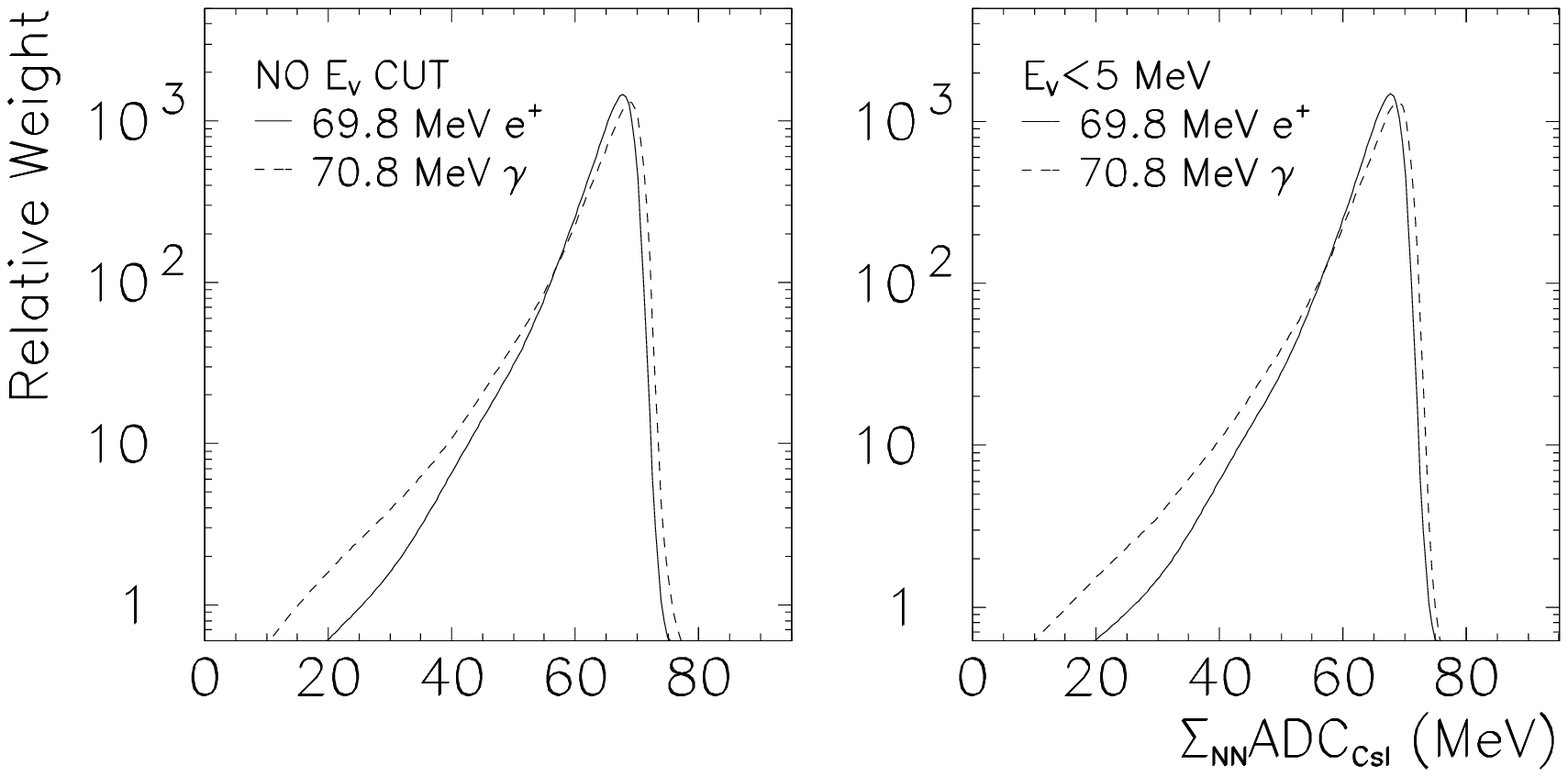,height=20cm}}
\vglue -8.5cm
\centerline{FIGURE 25}
\vspace*{\stretch{2}}
\clearpage

\end{document}